\documentclass[11pt,a4paper]{article}
\usepackage[a4paper,total={160mm,220mm}]{geometry}
\usepackage{amsmath,amssymb}
\usepackage{cite}
\usepackage{braket}
\usepackage[colorlinks=true, allcolors=blue, linktocpage=true]{hyperref}
\usepackage[capitalise]{cleveref}
\usepackage{caption}
\usepackage{subcaption}
\usepackage{graphicx}
\graphicspath{{figures/}}
\usepackage{tensor}

\newcommand{\abs}[1]{\left\lvert{#1}\right\rvert}
\newcommand{\absl}[1]{\lvert{#1}\rvert}
\newcommand{\pd}{\partial}

\newcommand{\bB}{\overline{B}}
\newcommand{\bvB}{\overline{\vec{B}}}
\newcommand{\bF}{\overline{F}}
\newcommand{\bJ}{\overline{j}}
\newcommand{\dF}{\delta F}
\newcommand{\dJ}{\delta j}
\newcommand{\md}{\mathrm{d}}
\newcommand{\me}{\mathrm{e}}
\newcommand{\htt}{h^{\mathrm{TT}}}
\newcommand{\hfn}{h^{\mathrm{FN}}}

\renewcommand{\vec}[1]{\mathbf{#1}}

\begin{document}
\numberwithin{equation}{section}
\title{
\vspace*{-0.5cm}{\scriptsize \mbox{}\hfill MITP-24-040}\\
\vspace{3.5cm}
\Large{\textbf{A Coordinate-Independent Formalism for Detecting High-Frequency Gravitational Waves}}
\vspace{0.5cm}}

\author{Wolfram Ratzinger,$^{1}$ Sebastian Schenk,$^{2}$ and Pedro Schwaller$^{2}$\\[2ex]
\small{\em $^{1}$Department of Particle Physics and Astrophysics, Weizmann Institute of Science,} \\
\small{\em Herzl Street 234, Rehovot, 7610001, Israel}\\[0.5ex]
\small{\em $^{2}$PRISMA$^{+}$ Cluster of Excellence \& Mainz Institute for Theoretical Physics,} \\
\small{\em Johannes Gutenberg-Universit\"at Mainz, 55099 Mainz, Germany}
\\[0.8ex]}

\date{}
\maketitle

\begin{abstract}
\noindent
In an external electric or magnetic field, a gravitational wave (GW) may be converted into electromagnetic radiation.
We present a coordinate-invariant framework to describe the GW signal in a detector that is based on this effect, such as cavities for axion searches.
In this framework, we pay special attention to the definition of manifestly coordinate-independent expressions for the electromagnetic fields that an external observer would detect.
A careful assessment of the detector's perceived motion allows us to treat both its mechanical and its electromagnetic response to the GW consistently.
We further introduce well-defined approximations for which this motion may be neglected, and hence provide suggestions on which coordinate frame is suitable to characterise the GW signal in practice.
We illustrate our findings in two examples, an infinitesimally thin rod and a spherical electromagnetic cavity.
\end{abstract}

\newpage
\tableofcontents

\section{Introduction}
\label{sec:introduction}

The detection of gravitational waves (GWs) by laser interferometers~\cite{LIGOScientific:2016aoc} and the growing evidence for a stochastic GW background suggested by pulsar timing arrays~\cite{NANOGrav:2023gor,EPTA:2023fyk,Reardon:2023gzh} provide crucial insights into the fundamental dynamics of our Universe.
For instance, these measurements put stringent bounds on modifications of gravity (see, e.g.,~\cite{LIGOScientific:2016lio,NANOGrav:2023hfp}).
It is therefore only natural that we seek to further improve the experimental sensitivity of GW measurements, as well as to expand the frequency range over which we can detect them.
Current and future GW detectors mostly address frequencies in the kHz range and below. 
On the other hand, at frequencies above 10~kHz there are no known GW sources of astrophysical origin, thereby providing an exceptional testbed for new physics beyond the Standard Model.
That said, a detection of a GW signal in this frequency regime would either imply the existence of exotic astrophysical objects, such as primordial black holes or boson stars, or open an entirely novel window into the early Universe~\cite{Aggarwal:2020olq}.

In addition, GWs are the only possible messenger reaching us from times long before big bang nucleosynthesis, with all other known particle species rapidly thermalising.
While established GW searches may already shed light on some of these early Universe intricacies, there is a plethora of scenarios that require an amplified sensitivity to GWs of higher frequencies.
In particular, causality implies that any source emitting GWs when the Universe's temperature was above $10^{10}$~GeV, assuming radiation domination at early times, inevitably leads to GW signals that are currently out of experimental reach.
GWs at higher frequencies may therefore probe a possible grand unification or even string scale, and could be sourced by phase transitions, topological defects or bosonic instabilities (see~\cite{Aggarwal:2020olq} for a recent review).

While laser interferometers may be able to push GW detection into the kHz to MHz range, any experimental sensitivity to GW frequencies beyond this regime requires radically new measurement techniques.
One possibility is the conversion of gravitational into electromagnetic waves in the presence of electromagnetic background fields due to the inverse Gertsenshtein effect~\cite{Gertsenshtein:1962}.
The technology required to harness this effect has seen a large boost in the last decades due to the inherent similarity with the conversion of axions in the vicinity of an external magnetic field~\cite{Raffelt:1987im}.
Indeed, there now is substantial experimental effort aimed at covering the axion parameter space (see, e.g.,~\cite{ADMX:2001dbg,Kahn:2016aff,Ouellet:2018beu,DMRadio:2022jfv,McAllister:2017lkb}), as well as proposals specifically designed for the detection of GWs~\cite{Ballantini:2005am,Gao:2023gph,Schmieden:2023fzn,Alesini:2023qed,Navarro:2023eii}.

Order-of-magnitude estimates regarding the sensitivity of such experiments to GWs date back to very early works~\cite{Lupanov:1967}.
From a theory point of view, it remains challenging to accurately capture this prediction using general relativity.
In this endeavour, one school of thought advertises the use of a ``proper detector frame" around the center of mass of the experimental apparatus, arguing that this frame is ``naturally connected" to the detector~\cite{Baroni:1984ptn,Callegari:1987ux}.
The idea of a distinguished frame seems, however, somewhat obscure in a geometric theory of gravity.
It has further been pointed out that if the GW wavelength and the size of the experiment are comparable, the detector appears to be moving in the proper detector frame while being at rest in transverse-traceless (TT) gauge.
This, on the other hand, suggests that the latter perhaps is a more suitable choice~\cite{Faraoni:1991bw}.
In fact, the situation is even more dire, as there is also disagreement on the correct definition of observables and hence both approaches may not even be equivalent.
For instance, there is no consensus on whether there may be an observable signal in detectors involving a static homogeneous magnetic background field with an incoming GW that is parallel to the field~\cite{DeLogi:1977qe,Herman:2020wao,Berlin:2021txa}.
Therefore, it seems that, even to this day, the question of how to consistently treat these scenarios has not been answered convincingly.
While various approaches are present in the literature, an established framework of how all of these are related to each other is still lacking.
We address this problem in this work.

The aim of this paper is to consistently describe the signal of GWs in an experiment that is based on their conversion into electromagnetic radiation.
We start by reviewing the theory of electrodynamics within general relativity in \cref{sec: electrodynamics and gravitational waves}, paying special attention to the definition of manifestly coordinate-independent expressions for observable electromagnetic fields and their associated boundary conditions.
We then introduce a perturbation scheme, focusing on the scenario where the metric tensor is decomposed into a flat Minkowski contribution and small fluctuations, the GWs.
Our scheme involves a choice of certain replacement rules for tensor quantities, from which we derive all equations of motion as well as the rules for infinitesimal coordinate (gauge) transformations.
We also demonstrate that the perturbation scheme is consistent in the sense that the equations of motion are manifestly invariant under these gauge transformations.

In our framework, the expressions for the observable electromagnetic fields and their boundary conditions are intricately related to the perceived frame-dependent motion of the detector.
In \cref{sec: mechanical deformations through gravitational waves} we therefore review the motion (and deformation) of elastic bodies due to an incoming GW.
In particular, we revisit the fact that, depending on the ratio between the GW's wavelength and the size of the detector, the experiment appears to be at rest in either the proper detector frame or TT gauge.
This allows us to introduce well-defined approximations in which this motion can be neglected when solving Maxwell's equations in the respective frame.
Along with these approximations we give a suitable frequency range where they are valid as well as parametric error estimates, such that we provide an answer to the naive question which frame is the most suitable to treat a specific problem.
We illustrate our findings in two detailed examples, an infinitesimally thin rod and a spherical electromagnetic cavity, in \cref{sec: thin rod,sec: spherical cavity}, respectively.
Finally, we summarise our results and conclude by outlining some practical approaches in \cref{sec: conclusions}.

\subsection*{Conventions}

Throughout this work, we use a mostly positive metric, $g_{\mu\nu}$, with $\eta_{\mu\nu}$ denoting the flat Minkowski metric, $\eta_{\mu \nu} = \mathrm{diag} (-1, 1, 1, 1)$.
Greek indices run from $\alpha = 0\ldots3$, and Latin indices from $a = 1\ldots3$.
Underlined indices, $\underline{\alpha}$, denote tetrad indices to be distinguished from the regular coordinate ones, $\alpha$.
A bold notation, $\mathbf{k}$, refers to the spatial components of a vector.
The Levi-Civita symbol is denoted by $\epsilon$, with normalisation $\epsilon_{0123}=\epsilon_{123}=1$.
The volume form, i.e.~the Levi-Civita tensor, is then given by $\Omega_{\mu\nu\rho\sigma} = \sqrt{-g} \epsilon_{\mu\nu\rho\sigma}$ in any right-handed coordinate system.
Finally, $\tau$ denotes proper time in all frames.

\section{Electrodynamics and Gravitational Waves}
\label{sec: electrodynamics and gravitational waves}

In general relativity, the dynamics of electromagnetic fields in vacuum are governed by Maxwell's equations in a curved spacetime,
\begin{align}
	\nabla_\nu F^{\mu \nu} &= j^{\mu} \, , \label{eq:MaxwellInhom} \\
	\pd_{[ \lambda} F_{\mu \nu]} &= 0 \label{eq:MaxwellHom} \, .
\end{align}
Here, $F^{\mu \nu}$ and $j^{\mu}$ denote the electromagnetic field strength tensor and current, respectively, and $\nabla_\mu$ is the covariant derivative along the $\mu$-direction.
Furthermore, the square brackets indicate a sum over all cyclic permutations.
While Maxwell's equations are manifestly invariant under a general coordinate transformation, the tensor quantities $F^{\mu \nu}$ and $j^{\mu}$ by themselves are not.
Their components therefore do not correspond to any physical observable.
Instead, one needs to carefully establish the electric and magnetic fields an observer would measure in the presence of a gravitational field.

\subsection{Observable electromagnetic fields}
\label{sec: observable electromagnetic fields}

To determine the electromagnetic fields measured by an external observer travelling along a given worldline, we closely follow Chapter~6 of Ref.~\cite{Misner:1973prb}, and associate an infinitesimal proper coordinate system to the observer's worldline.
That is, we choose a set of four independent vector fields, an orthonormal tetrad $e_{\underline{\alpha}}(\tau)$, with coefficients $e^\mu_{\underline{\alpha}}$ in a given coordinate basis.
By definition, the zeroth field coincides with the observer's four-velocity, $e^\mu_{\underline{0}} = u^\mu$, while the components in general obey the equations of motion~\cite{Misner:1973prb}
\begin{equation}
	\frac{\md}{\md \tau} e^\mu_{\underline{\alpha}} + \Gamma^\mu_{\nu \lambda} u^\nu e^\lambda_{\underline{\alpha}} = \left( a_\nu u^\mu - a^\mu u_\nu \right) e^\nu_{\underline{\alpha}} + u^\lambda \omega^\rho \tensor{\Omega}{_\lambda _\rho _\nu ^\mu} e^\nu_{\underline{\alpha}} \, .
\label{eq:EOMTetrad}
\end{equation}
Here, $a^\mu$ and $\omega^\mu$ are the observer's acceleration and rotation, respectively.
The acceleration and rotation are both orthogonal to the observer's four-velocity and, more importantly, are both measurable quantities.
They can be assessed as follows.
Consider, for instance, the force that is needed to keep a test particle of mass $m$ on a trajectory following the observer's worldline, $K^\mu = m a^\mu$.
This force can be measured relative to the coordinate system defined by the tetrad, such that the spatial components of the observed acceleration become $a_{\underline{i}} = a_\mu e^\mu_{\underline{i}}$.
Similarly, the relative rotation of the spatial components of the tetrad can be compared to a non-rotating tetrad, $e^{\prime}_{\underline{i}}$, realised by a gyroscope.
For instance, at a time where both tetrads coincide, they are related by~\cite{Misner:1973prb}
\begin{equation}
	\frac{\md}{\md \tau} \left( e_{\underline{a}}^{\mu} - {e^{\prime}}_{\underline{a}}^{\mu} \right) = \tensor{\epsilon}{_{\underline{a}} ^{\underline{b}} ^{\underline{c}}} \omega_{\underline{b}} e_{\underline{c}}^{\mu} \, ,
\end{equation}
with the relative rate of rotation $\omega_{\underline{i}}=\omega_\mu e^\mu_{\underline{i}}$. 

These examples can be generalised to an environment involving electromagnetic fields.
For instance, let us consider a scenario where two test masses are forced to follow the observer's worldline.
If one of the test masses carries an electric charge $q$, the relative difference in forces acting on both is given by $\Delta K^\mu = q \tensor{F}{^\mu _\nu} u^\nu$.
At the same time, this force should correspond to $\Delta K_{\underline{a}} = q E_{\underline{a}}$ and hence the observed electric field reads
\begin{equation}
	E_{\underline{a}} = F_{\mu \nu} e^\mu_{\underline{a}} u^\nu \, .
\label{eq:Eobs}
\end{equation}
Similarly, if we introduce a relative velocity between two test masses of equal charge, $u^\mu \to u^\mu + e^\mu_{\underline{a}} v^{\underline{a}}$, we find for the relative force between both $\Delta K^\mu = q \tensor{F}{^\mu _\nu} e^\nu_{\underline{a}} v^{\underline{a}}$.
Again, this should correspond to the Lorentz force $\Delta K_{\underline{a}} = q \epsilon_{\underline{abc}} v^{\underline{b}} B^{\underline{c}}$, such that the observed magnetic field is given by
\begin{equation}
	B^{\underline{a}} = \frac{1}{2} \epsilon^{\underline{abc}} F_{\mu \nu} e^\mu_{\underline{b}} e^\nu_{\underline{c}} \, .
\label{eq:Bobs}
\end{equation}
We remark that in this tetrad formalism there are no electric and magnetic fields that are defined globally.
Instead, they are only defined for one observer at a time relative to their local coordinates given by the tetrad.
This is in contrast to previous works where, e.g., the components $F_{i 0}$ in Fermi normal (FN) coordinates are interpreted as global electric fields, arguing that this be a ``natural" choice~\cite{Baroni:1984ptn, Baroni:1985gwt}.
That said, it is of course legitimate to define these quantities in any coordinate system, because the homogeneous part of  Maxwell's equations~\eqref{eq:MaxwellHom} is the same as in flat spacetime.
Nevertheless, they are merely components of a tensor and have therefore, by themselves, no physical meaning.
We will later comment on scenarios where this choice can still be a useful approximation in an experimental setting.

\bigskip

Similar to the electromagnetic fields, we can treat their boundary conditions in a coordinate-invariant way.
These uniquely determine the solutions to Maxwell's equations.
Let us, for instance, consider an observer attached to the surface of a conducting material.
By definition, inside the conductor, the electric field measured by the observer vanishes.
We can then assign an orthonormal tetrad along the observer's worldline, such that the zeroth component aligns with the observer's four-velocity, $e_{\underline{0}}^{\mu} = u^{\mu}$, and both $e_{\underline{1}}^{\mu}$ and $e_{\underline{2}}^{\mu}$ are tangential to the surface.
In the presence of an electromagnetic field, the force tangential to the surface acting on a charge $q$ following the observer is $q F_{\mu \nu} e_{\underline{1}, \underline{2}}^{\mu} u^{\nu}$.
However, in the limit of vanishing resistance at the surface of the conductor, i.e.~at the interface between the conducting material and vacuum, this force would lead to a diverging current, such that the boundary condition for an ideal conductor in this framework has to be~\cite{Rawson-Harris:1972nfp}
\begin{equation}
	F_{\mu \nu} e_{\underline{1}}^{\mu} u^\nu = 0 = F_{\mu \nu} e_{\underline{2}}^{\mu} u^{\nu} \, .
\label{eq:BCCoordinateIndependent}
\end{equation}
Indeed, this is the equivalent of the well-known boundary condition $\vec{E}_{\parallel} = 0$. 
In a similar manner one can find the equivalent for the relations involving the magnetic and orthogonal electric field. 
These have been derived for an interface between arbitrary media in \cite{Rawson-Harris:1972nfp}.
For the special case of a conductor, we provide a detailed discussion of these relations in \cref{app:ElasticConductor}.

\bigskip

In summary, we have defined a framework to properly characterise the observable electromagnetic fields that an external observer would detect in the presence of a gravitational field.
Let us now apply these ideas to the interaction between electromagnetism and GWs.

\subsection{A perturbation scheme for weak gravitational fields}

We now focus on the dynamics of electromagnetic fields in the presence of a GW, i.e.~a weak gravitational perturbation.
Typically, the latter is parametrised by a small fluctuation around a flat background metric,
\begin{equation}
	g_{\mu \nu} = \eta_{\mu \nu} + h_{\mu \nu} \, ,
\end{equation}
with $\absl{h_{\mu \nu}} \ll 1$.
Here, $\eta_{\mu \nu}$ is the flat Minkowski background metric and $h_{\mu \nu}$ denotes the GW.

Similarly, we expand the electromagnetic field strength tensor and current,
\begin{align}
	F_{\mu \nu} &= \bF_{\mu \nu} + \dF_{\mu \nu} \, , \label{eq:PerturbationSchemeF} \\
	j^{\mu} &= \bJ^{\mu} + \dJ^{\mu} \, .
\end{align}
Here, we assume that typical strains asserted by a GW are small such that quantities preceded by $\delta$ are of order $\mathcal{O}(h)$, while all leading-order quantities are denoted by a bar.
That is, we only consider contributions which are at most of order $\mathcal{O}(h)$, and neglect any higher-order terms. 

We remark that our ansatz for the expansion \emph{defines} a perturbation scheme.
This is because the indices of the unperturbed quantities are raised and lowered using the metric tensor $g_{\mu \nu}$, while the indices of all quantities in the perturbative expansion are raised and lowered using the flat background metric tensor $\eta_{\mu \nu}$.
Therefore, the choice of perturbation associated to $F_{\mu \nu}$ in \cref{eq:PerturbationSchemeF} instead of its contravariant counterpart $F^{\mu \nu}$ will lead to different equations of motion governing the dynamics.\footnote{See also the discussion around the different contributions to the effective current~\eqref{eq:jeff}.}
However, when applied consistently, any choice of scheme will lead to the same physical observables (see, e.g., \cite{Maggiore:2007ulw}).
We provide more details on this in \cref{app:PerturbationScheme}.

Having defined a perturbation scheme, we can express Maxwell's equations order by order in the weak-field expansion.
In our scheme, the homogeneous equations~\eqref{eq:MaxwellHom} remain trivial, i.e.~they are satisfied by both $\bF_{\mu \nu}$ and $\dF_{\mu \nu}$ separately, which may not be the case for a different choice of scheme.
The perturbative treatment of the inhomogeneous equations~\eqref{eq:MaxwellInhom} is more involved.
At the leading order, we find the trivial relation
\begin{equation}
	\pd_\nu \bF^{\mu \nu} = \bJ^{\mu} \, .
\label{eq:EOMLeading}
\end{equation}
At the first order of the perturbative expansion the fluctuations satisfy
\begin{equation}
	\pd_\nu \dF^{\mu \nu} = \dJ^{\mu} + j_{\mathrm{eff}}^{\mu} \, ,
\label{eq:EOMFirstOrder}
\end{equation}
with an effective current given by
\begin{equation}
	j_{\mathrm{eff}}^{\mu} = - \frac{1}{2} \left( \pd_\lambda h \right) \bF^{\mu \lambda}  + \pd_\nu \left( \tensor{h}{^\mu_\lambda} \bF^{\lambda \nu} + \tensor{h}{^\nu_\lambda} \bF^{\mu \lambda} \right) \, .
\label{eq:jeff}
\end{equation}
Here, $h$ denotes the trace of the metric fluctuations, $h = \tensor{h}{^\mu _\mu}$.
The effective current clearly illustrates that a GW sources electromagnetic fields in classical electrodynamics.
It has two distinct contributions.
The first one arises from the Levi-Civita connection of the metric, where to first order in the perturbation scheme $\nabla_{\nu} F^{\mu \nu} = \pd_{\nu} F^{\mu \nu} + \Gamma_{\nu \lambda}^{\nu} F^{\mu \lambda} = \pd_{\nu} F^{\mu \nu} + \frac{1}{2} \pd_{\lambda} h F^{\mu \lambda}$.
The second contribution to the effective current is due the fluctuations of $F^{\mu \nu}$ in our choice of perturbation scheme.
This can be seen by lowering the indices of the contravariant tensor first, and then applying the perturbative expansion~\eqref{eq:PerturbationSchemeF}.
This yields $F^{\mu \nu} = g^{\mu \lambda}g^{\nu \rho} F_{\lambda \rho} = \bF^{\mu \nu} + \delta F^{\mu \nu} - \tensor{h}{^\mu _\lambda} \bF^{\lambda \nu} - \tensor{h}{^\nu _\rho} \bF^{\mu \rho}$, where we have used the perturbative relation $g^{\mu \nu} = \eta^{\mu \nu} - h^{\mu \nu}$.
Clearly, this does not only apply to the electromagnetic field strength tensor and current, but also to the observer's position and tetrad.
We define these as
\begin{align}
	x^{\mu} &= \overline{x}^{\mu} + \delta x^{\mu} \, , \\
	e_{\underline{\alpha}}^{\mu} &= \overline{e}_{\underline{\alpha}}^{\mu} + \delta e_{\underline{\alpha}}^{\mu} \, .
\end{align}
Similar to our earlier discussion, this choice at the same time implies that the covariant equivalents of these vectors pick up additional terms proportional to the metric perturbation.

\bigskip

Let us close this discussion with a few remarks on coordinate transformations in this weak-field expansion.
Crucially, our approach needs to be invariant under infinitesimal coordinate transformations,
\begin{equation}
	x^{\prime \mu} = x^{\mu} + \xi^{\mu} (x) \, ,
\end{equation}
where the spacetime-dependent shift $\xi^{\mu}$ is considered to be sufficiently small, $\pd_\mu \xi^{\mu} \sim \mathcal{O}(h)$, to allow for a consistent perturbative treatment.
From the general transformation properties of tensors, we find that the leading-order terms $\bF^{\mu \nu}$ and $\bJ^{\nu}$ do not transform under coordinate shifts, while the fluctuations satisfy
\begin{align}
	h^{\prime}_{\mu \nu} &= h_{\mu \nu} - \pd_\mu \xi_\nu - \pd_\nu \xi_\mu \, , \\
	\dF^{\prime}_{\mu \nu} &= \dF_{\mu \nu} - \xi^{\lambda} \pd_{\lambda} \bF_{\mu \nu} - \bF_{\lambda \nu} \pd_\mu \xi^{\lambda} - \bF_{\mu \lambda} \pd_\nu \xi^{\lambda} \, , \label{eq:ICTElectromagneticStressTensor} \\
	\dJ^{\prime \mu} &= \dJ^{\mu} - \xi^{\lambda} \pd_{\lambda} \bJ^{\mu} + \bJ^{\lambda} \pd_{\lambda} \xi^\mu \, .
\end{align}
Here, primed quantities correspond to the new coordinate frame.
With these transformation laws, one can check explicitly that the equations of motion are invariant under infinitesimal coordinate transformations.
Indeed this is by construction, starting from Maxwell's equations~\eqref{eq:MaxwellInhom} and~\eqref{eq:MaxwellHom} and consistently applying the perturbation scheme.
Strictly speaking, anything that cannot be formulated in a fully covariant manner will inevitably lead to an answer that is ill-defined within the context of general relativity.
Clearly, this means the electric and magnetic field as components of the electromagnetic field strength tensor, e.g., $E_a = F_{0a}$, are not invariant under coordinate transformations.
On the other hand, crucially, using the transformation properties of the observer's position and tetrad under infinitesimal coordinate transformations,
\begin{align}
    \delta {x^{\prime}}^\mu &= \delta {x}^\mu+\xi^\mu \, , \\
	\delta {e^{\prime}}_{\underline{\alpha}}^{\mu} &= \delta e_{\underline{\alpha}}^{\mu} + \overline{e}_{\underline{\alpha}}^{\nu} \pd_{\nu} \xi^{\mu} \, ,
\end{align}
the observable fields, e.g., $E_{\underline{a}} = F_{\mu \nu} e_{\underline{a}}^{\mu} u^{\nu}$ are indeed invariant, and therefore well defined.
We illustrate this more explicitly and provide a detailed summary of our perturbation scheme in \cref{app:PerturbationScheme}.
In practice, as these observables are manifestly coordinate-independent, we can perform their calculation in any coordinate frame.
Some frames, however, may be more suitable than others in certain scenarios, as we will discuss below.

\subsection{Coordinate frames in action}
\label{sec:coordinate_frames}

Our framework, so far, is completely independent of a specific choice of coordinates.
In practice, however, one has to perform calculations of signal estimates using a suitable coordinate frame.
Let us briefly review two of the most commonly used coordinate frames.

\paragraph{Transverse-traceless gauge}

In the so-called transverse-traceless (TT) gauge, the metric fluctuations obey
\begin{equation}
	\pd_\mu \partial^\mu \htt_{\alpha \beta} = 0 \, , \enspace
	\htt_{0 \mu} = 0 \, , \enspace
	\partial^i \htt_{ij} = 0 \, , \enspace
	\tensor{{\htt}}{^i _i} = 0 \, .
\end{equation}
In this frame, a monochromatic GW with wave vector $\vec{k}$ and frequency $\omega = \absl{\vec{k}}$ is given by
\begin{equation}
	\htt_{ij} = \left( A^{+}_{ij} + A^{\times}_{ij} \right) \me^{i \left( \omega t - \vec{k} \cdot \vec{x} \right)} \, .
\end{equation}
Here, the linear polarisation coefficients are $A^{+}_{ij} = A^{+} ( \hat{e}^1_i \hat{e}^1_j - \hat{e}^2_i \hat{e}^2_j )$ and $A^{\times}_{ij} = A^{\times} ( \hat{e}^1_i \hat{e}^2_j + \hat{e}^2_i \hat{e}^1_j )$, where $\vec{\hat{e}}^1$ and $\vec{\hat{e}}^2$ are chosen such that they form an orthonormal basis together with the unit vector in the propagation direction of the GW, $\vec{\hat{k}} = \vec{k} / \absl{\vec{k}}$.
A decisive feature of TT coordinates is that they are synchronous, $\htt_{0 \mu} = 0$.
As we will later see, this property makes them an appealing choice when treating mechanical problems, as the motion of freely-falling masses that are at rest in the unperturbed system appear unaffected in this frame.

\paragraph{Fermi-normal coordinates}

Fermi-normal (FN) coordinates, sometimes called the proper detector frame, are constructed by starting from the worldline of an observer and then extending spacelike vectors orthogonal to the observer's velocity into geodesics, thereby spanning a worldtube around the observer's worldline.
Starting from a freely-falling and freely-rotating observer at the origin in the spacetime defined in the previous section, the coordinate transformation from TT to FN coordinates is given by~\cite{Faraoni:1991bw, Marzlin:1994ia}
\begin{align}
	t^{\mathrm{FN}} &= t - \frac{i}{2} \omega \htt_{ij} (t) x^i x^j \mathcal{F} ( \vec{k} \cdot \vec{x} ) \, , \\
	x^{\mathrm{FN}}_i &= x_i + \htt_{ij} (t) x^j \left( \frac{1}{2} + i \vec{k} \cdot \vec{x} \mathcal{F} (\vec{k} \cdot \vec{x}) \right) - \frac{i}{2} k_i \htt_{mn} (t) x^m x^n \mathcal{F} (\vec{k} \cdot \vec{x}) \, .
\end{align}
Here, $\vec{k}$ is the wave vector of the GW and we have defined $\mathcal{F} (\xi) = (\me^{-i \xi} - 1 + i \xi)/\xi^2 \approx -1/2 + \mathcal{O} (\xi)$.
Furthermore, we have used the notation $\htt_{ij}(t) \equiv \htt_{ij}(t, \vec{x})\rvert_{\vec{x} = 0}$ to denote the metric fluctuation in TT gauge at the observer's position.
Following~\cite{Berlin:2021txa, Domcke:2022rgu}, the metric fluctuations in the proper detector frame become (see also~\cite{1982NCimB..71...37F, Marzlin:1994ia, Rakhmanov:2014noa})
\begin{align}
	\hfn_{00} &= \omega^2 \mathcal{F}(\vec{k} \cdot \vec{x}) \vec{b} \cdot \vec{x} \, , \\
	\hfn_{0i} &= \frac{\omega^2}{2} \left[ \mathcal{F} (\vec{k} \cdot \vec{x}) + i \mathcal{F}^{\prime} (\vec{k} \cdot \vec{x}) \right] \left( \vec{\hat{k}} \cdot \vec{x} b_i - \vec{b} \cdot \vec{x} \hat{k}_i \right) \, , \\
	\hfn_{ij} &= i \omega^2 \mathcal{F}^{\prime} (\vec{k} \cdot \vec{x}) \left( \abs{\vec{x}}^2 \htt_{ij} (t) + \vec{b} \cdot \vec{x} \delta_{ij} - b_i r_j - b_j r_i \right) \, ,
\end{align}
where we have defined $b_j = x_i \htt_{ij} \rvert_{\vec{x} = 0}$.
Here, $\mathcal{F}^{\prime}$ denotes the derivative of $\mathcal{F}$ with respect to its argument, $\mathcal{F}^{\prime}(\xi) \approx i/6 + \mathcal{O}(\xi)$.
Clearly, the corrections to the flat Minkowski metric are of order $\mathcal{O} \left(\omega^2 x^2\right)$ and higher.
Physically, they encode the lowest order gravitational strain that can be measured by an external observer~\cite{Manasse:1963zz}.
This property also makes the FN frame of interest for mechanical problems where a small rigid body is able to withstand this strain and appears to be at rest, approximately.

FN coordinates are a somewhat natural choice when considering an experimental apparatus that is much smaller than the wavelength of the GW, $\omega L \ll 1$, where $L$ is the characteristic length scale of the detector.
In this regime, a \textbf{long-wavelength approximation} is typically employed.
This can be implemented by performing a series expansion of each frequency-dependent quantity in powers of $\omega L$, and then retaining only the leading-order terms.
For instance, any correction to the metric perturbation beyond the quadratic order is neglected by truncating $\mathcal{F}$ to $\mathcal{F} \approx -1/2$ and $\mathcal{F}^{\prime} \approx i / 6$.
In general, a similar truncated series expansion has to be performed for each object in the long-wavelength approximation.

We finally remark that, by construction, the observer's tetrad that is expanded into FN coordinates has trivial components, $e_{\underline{\alpha}}^\mu=\delta_{\underline{\alpha}}^\mu$, where the Kronecker-$\delta$ now mixes coordinate and tetrad indices.
Therefore, our definitions of the observed electromagnetic fields align with schematically identifying, e.g., $E_{\underline{i}} (\tau) = F_{\underline{i} 0} (\tau) \rvert_{\vec{x} = 0}$, where $F_{\mu \nu}$ are now the components of the electromagnetic field strength tensor in the FN basis.
Indeed, at the origin this relation is an exact equality.
For a small displacement from it, one may still be able to interpret the electric field defined in this way as the quantity that is measured by an observer following the original one at a fixed physical distance.
This, however, is only an approximation that deteriorates at larger distances.

\section{Detector Deformations through Gravitational Waves}
\label{sec: mechanical deformations through gravitational waves}

The expressions for the observable electromagnetic fields and their associated boundary conditions, given in \cref{eq:Eobs,eq:BCCoordinateIndependent}, in general correlate the electromagnetic and mechanical response of an experimental apparatus to an incoming GW.
More precisely, this relation is built upon the observer's four-velocity, $u^{\mu}$.
Clearly, an incoming GW will perturb the latter through a mechanical deformation of the apparatus, $\delta \vec{x}$, as schematically it is $\delta \vec{u} = \delta \dot{\vec{x}}$.
Focusing on elastic solids, the dynamics of these mechanical deformations in weak gravitational fields are governed by~\cite{Papapetrou1972, Hudelist:2022ixo}
\begin{equation}
	\rho \left[ \pd_t^2 \delta x^i + \frac{1}{2} \left( 2 \pd_t \tensor{h}{_0 ^i} - \partial^i h_{00} \right) \right] = \pd_j \sigma^{ij} \, ,
\label{eq:MechanicsEOM}
\end{equation}
together with the boundary condition
\begin{equation}
	\left. \sigma^{ij} n_j \right\rvert_{\partial V} = 0 \, .
\label{eq:MechanicsBC}
\end{equation}
Here, the deformation is treated as a function of time and space inside a volume $V$ occupied by the body that is assumed to be at rest in the unperturbed system, and $\vec{n}$ is a vector normal to the body's surface $\partial V$.
Furthermore, $\sigma^{ij}$ are the components of the Cauchy stress tensor~\cite{Hudelist:2022ixo},
\begin{equation}
	\sigma^{ij} = \lambda \delta^{ij} \tensor{e}{^k _k} + 2 \mu e^{ij} \enspace \mathrm{with} \enspace e^{ij} = \frac{1}{2} \left( \partial^i \delta x^j + \partial^j \delta x^i + h^{ij} \right) \, .
\label{eq:StressTensor}
\end{equation}
For simplicity, we have assumed that the material is homogeneous and isotropic, such that $\lambda$ and $\mu$ are scalar Lam\'e parameters.

Indeed, it is helpful to investigate the mechanical response in TT and FN coordinates specifically, because it can be safely neglected in certain regimes.
This, in turn, may significantly simplify the characterisation of the mechanical detector response to an incoming GW.

\paragraph{TT coordinates}

In the TT frame, the bulk contribution of the metric fluctuations vanish, such that the mechanical deformations satisfy a free wave equation,
\begin{equation}
	\rho \pd_t^2 \delta {x^{\mathrm{TT}}}^i = \pd_j {\sigma^{\mathrm{TT}}}^{ij} \, .
\end{equation}
In this case, the metric fluctuations only enter the boundary condition~\eqref{eq:MechanicsBC} via the Cauchy stress tensor
\begin{equation}
	{\sigma^{\mathrm{TT}}}^{ij} = \lambda \delta^{ij} \partial^k \delta x^{\mathrm{TT}}_k + \mu \left( \partial^i \delta {x^{\mathrm{TT}}}^j + \partial^j \delta {x^{\mathrm{TT}}}^i + {\htt}^{ij} \right) \, .
\end{equation}
Therefore, in the limit of vanishing sound velocity, $v_s^2 \sim \lambda / \rho \sim \mu / \rho \ll 1$, the body appears to be at rest in the TT frame, $\pd_t^2 \delta x^{\mathrm{TT}}_i = 0$.
More precisely, one finds that in the regime $\omega L \lesssim v_s$ the mechanical deformations follow the naive estimate $\delta x \sim \mathcal{O}(h L)$, where $L$ is the characteristic length scale of the detector.
On the other hand, in the regime $\omega L \gtrsim v_s$ the deformations are suppressed by the small sound velocity, $\delta x \sim \mathcal{O}(hL v_s / (\omega L))$.
Carefully note that this estimate, however, neglects mechanical resonances that may appear for $\omega L \gtrsim v_s$.

As the sound velocity in elastic solids is typically small, $v_s \lesssim 10^{-5}$, setting $\delta \vec{x}^{\mathrm{TT}} = 0$ can provide a suitable approximation over a wide range of frequencies.
This approximation is often referred to as the \textbf{free-falling limit}, because it appears as if the apparatus is made of non-interacting freely-falling particles.
Intuitively, the low velocity of sound physically prevents the information of the deformation to spread across the body.
Therefore, the body cannot possibly react to the incoming GW~\cite{Bringmann:2023gba}.

\paragraph{FN coordinates at long wavelengths}

As we have pointed out previously, FN coordinates are a somewhat natural choice when considering an experimental apparatus that is much smaller than the wavelength of the incoming GW, $\omega L \ll 1$.
In FN coordinates, with the coordinate origin at the center of mass of the experiment and performing the \textbf{long-wavelength approximation} introduced in \cref{sec:coordinate_frames}, the mechanical deformations are driven by a bulk force,
\begin{equation}
	\rho \left[ \pd_t^2 \delta {x^{\mathrm{FN}}}^i - \frac{1}{2} \partial^i \hfn_{00} \right] = \pd_j {\sigma^{\mathrm{FN}}}^{ij} \, ,
\label{eq:MechanicsEOMFN}
\end{equation}
and obey a trivial boundary condition that does not explicitly involve the metric perturbation,
\begin{equation}
	\left. \sigma^{\mathrm{FN}ij} n_j \right\rvert_{\partial V} = 0 \enspace \mathrm{with} \enspace {\sigma^{\mathrm{FN}}}^{ij} = \lambda \delta^{ij} \partial^k \delta x^{\mathrm{FN}}_k + \mu \left( \partial^i {\delta x^{\mathrm{FN}}}^j + \partial^j {\delta x^{\mathrm{FN}}}^i \right) \, .
\end{equation}
Here, we neglect terms of order $\mathcal{O} \left(\omega^2 L^2\right)$ and higher.
In contrast to the situation in the TT frame, the body appears to follow deformations of order $\delta x \sim \mathcal{O}(hL)$ in the high-frequency regime $\omega L \gtrsim v_s$.
On the other hand, these deformations are suppressed, $\delta x \sim \mathcal{O} \left(hL \omega^2 L ^2 / v_s^2\right)$, at low frequencies, $\omega L \lesssim v_s$.
Naively, if the frequency of the incoming GW is low enough, the body can relax back into its original shape, preserving the proper distance of all mass points from the center of mass.
That is, by construction of FN coordinates, the body appears to be at rest.
Therefore, setting $\delta \vec{x}^{\mathrm{FN}}=0$ may provide a useful approximation in this regime.
This is often referred to as the \textbf{rigid limit.}

\section{The Thin Rod}
\label{sec: thin rod}

To illustrate our formalism, let us consider a toy example of an infinitesimally thin rod of length~$L$.
We assume it to be oriented along the $x$-axis, extending from $-L/2$ to $L/2$ in an unperturbed system.
We further consider the observer to be mounted to one end of the rod, at $x=L/2$, and an orthonormal tetrad $e^\mu_{\underline{\alpha}}$ along its worldline that characterises the direction the sensors are pointing to.
For simplicity, we assume that all sensors are able to freely rotate, such that $\omega^{\mu} = 0$.

The rod is placed in a homogeneous magnetic field, $\bB$, which is pointing along the $z$-axis such that the only nonvanishing components of the electromagnetic field strength tensor are
\begin{equation}
	\bF_{21} = - \bF_{12} = \bB \, .
\end{equation}
Let us further consider a monochromatic GW propagating in the $z$-direction, parallel to the magnetic field.
Assuming it has an $A^{+}$ polarisation that is aligned with the $x$- and $y$-axis, the components of the metric perturbation in the TT frame read
\begin{equation}
	\htt_{11} = - \htt_{22} = A^{+} \me^{i \omega \left( t - z \right)} \, ,
\end{equation}
with all other components being zero.
In this scenario, the effective current~\eqref{eq:jeff} vanishes in TT gauge, $j^{\mu}_{\mathrm{eff}} = 0$.
Assuming that there are no other currents present, $\bJ^{\mu} = \dJ^{\mu} = 0$, the equations of motion for the fluctuations of the electromagnetic field strength simplify to
\begin{equation}
	\pd_\nu \delta F^{\mu \nu} = 0 \, .
\end{equation}
The solution to these equations is uniquely specified by a set of appropriate boundary conditions, e.g., implementing shielding against electromagnetic radiation from outside the detector.
For simplicity, we assume that there exists a set of boundary conditions that imposes a vanishing field strength altogether,
\begin{equation}
	\delta F^{\mu \nu} = 0 \, .
\end{equation}
Using the infinitesimal coordinate transformations~\eqref{eq:ICTElectromagneticStressTensor}, these fluctuations can be translated into FN coordinates, leading to a nonvanishing component of the field strength tensor,
\begin{equation}
	\delta F^{\mathrm{FN}}_{02} = \frac{i}{2} \omega x A^{+} \bB \me^{i \omega t} \, .
\label{eq:F_FN_rod}
\end{equation}
That is, in the proper detector frame one finds a nonvanishing electric field pointing in the $y$-direction.
Naively regarding this component, $\delta F^{\mathrm{FN}}_{02}$, as the measured electric field leads to an apparent contradiction.
It has been pointed out in~\cite{Berlin:2021txa} that this contradiction is resolved if one specifies whether the observer is at rest in TT or FN coordinates, hence corresponding to a freely-falling or rigidly-mounted observer, respectively.
However, going beyond earlier works, in our example the distinction between these regimes is in fact traced back to the finite sound velocity of the rod's material, and the intermediate regime can be treated consistently as we will demonstrate below.

\subsection{The observable electric field}

Let us clarify the situation within our coordinate-invariant framework, and find the electric field that is measured by the observer.
Applying our perturbation scheme to \cref{eq:Eobs}, we expand the observed field in $y$-direction to first order in the metric fluctuation,
\begin{equation}
	E_{\underline{2}} = \delta x^{\lambda} \left(\pd_{\lambda} \bF_{\mu \nu}\right) \overline{e}_{\underline{2}}^{\mu} \overline{u}^{\nu}
					+ \delta F_{\mu \nu} \overline{e}_{\underline{2}}^{\mu} \overline{u}^{\nu}
					+ \bF_{\mu \nu} \delta e_{\underline{2}}^{\mu} \overline{u}^{\nu}
					+ \bF_{\mu \nu} \overline{e}_{\underline{2}}^{\mu} \delta u^{\nu} \, .
\label{eq:observed_E_field}
\end{equation}
In principle, to find the observed field, we need to determine the perturbations $\delta x^{\lambda}$, $\delta F_{\mu \nu}$, $\delta e_{\underline{2}}^{\mu}$, and $\delta u^{\nu}$.
We note, however, that in practice several terms of this expression vanish.
First of all, the background field is static and homogeneous, such that the first term does not contribute, $\pd_{\lambda} \bF_{\mu \nu} = 0$.
Similarly, in the TT frame, we have already established that $\dF_{\mu \nu} = 0$, such that the second term vanishes, too.
Furthermore, according to our choice of perturbation scheme, the background values of the observer's tetrad are trivial, $\overline{e}^{\mu}_{\underline{\alpha}} = \delta^{\mu}_{\underline{\alpha}}$, and therefore $\overline{u}^{\nu} = \delta^{\nu}_{\underline{0}}$.
This again does not give a contribution to the observable electric field, because the electric field components of the background vanish, $\bF_{0\mu} = 0$.
Therefore, in TT gauge, the only nonvanishing term is the last one involving the observer's four-velocity perturbation.
Let us first establish the latter in the TT frame before commenting on the transformation into FN coordinates.

\begin{figure}[t]
	\centering
	\includegraphics[width=0.7\textwidth]{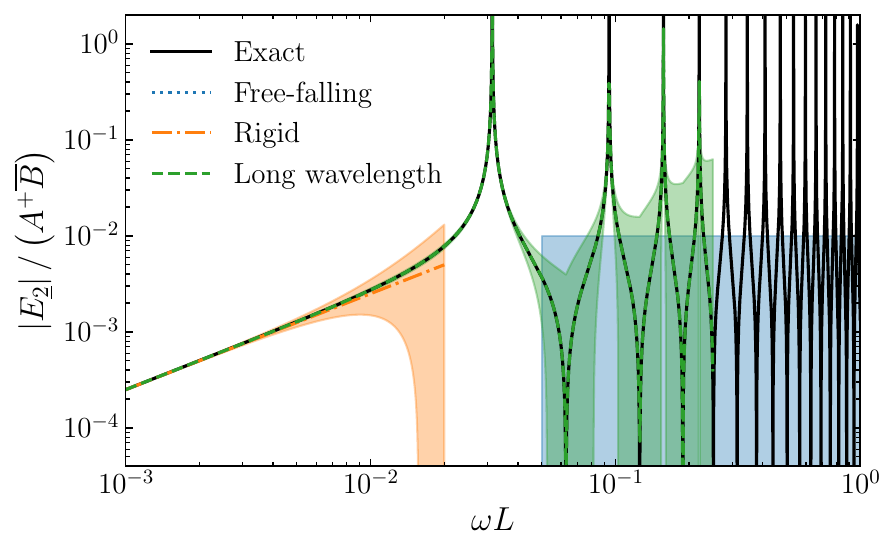}
	\caption{Amplitude of the observed electric field component in $y$-direction measured by an observer attached to the end of an infinitesimally thin rod of length $L$, as a function of $\omega L$. The rod is placed in a homogeneous magnetic background field $\bB$ pointing in the $z$-direction, while a GW with amplitude $A^{+}$ travelling parallel to the magnetic field perturbs the system. The sound velocity is chosen to be $v_s = 10^{-2}$.}
	\label{fig:e_rod}
\end{figure}

Mechanically, an infinitesimally thin rod can only support stress along its orientation direction, i.e.~the only nonvanishing component of the stress tensor is $\sigma^{11}$.
In the TT frame, the mechanical deformations in \cref{eq:MechanicsEOM} therefore satisfy~\cite{Hudelist:2022ixo}
\begin{equation}
	\pd_t^2 \delta x^1 = v_s^2 \pd_x^2 \delta x^1 \, , \enspace \pd_t^2 \delta x^2 = 0 \, , \enspace \pd_t^2 \delta x^3 = 0 \, ,
\label{eq:eom_rod}
\end{equation}
where we have defined the longitudinal sound velocity as $v_s = \sqrt{\mu (3 \lambda + 2\mu) / (\rho (\lambda + \mu))}$.
Furthermore, the mechanical deformations satisfy the boundary conditions~\cite{Hudelist:2022ixo}
\begin{equation}
	\left. \left( 2 \pd_x \delta x^1 + \htt_{11} \right) \right\rvert_{x=\pm \frac{L}{2}} = 0 \, .
\end{equation}
Let us now assume that the initial conditions are such that $\delta x^2 = \delta x^3 = 0$.
Indeed, an infinitesimally thin rod cannot provide resistance against deformations in the $y$- and $z$-direction, i.e.~the speed of sound for transverse waves vanishes.
The solution to the equations of motion~\eqref{eq:eom_rod}, subject to the above boundary condition, then reads
\begin{equation}
	\delta x^1 = - \frac{A^{+} L}{4 \chi \cos \chi} \sin \left( \frac{\omega x}{v_s} \right) \me^{i \omega t} \, ,
\end{equation}
where we have defined $\chi = \omega L / (2 v_s)$.
That said, the only nonvanishing component of the observer's four-velocity in the TT frame due to the mechanical deformation by the incoming GW is
\begin{equation}
	\delta u^1 = \left. \pd_t \delta x^1 \right\rvert_{x=\frac{L}{2}} = - \frac{i}{2} v_s A^{+} \tan \chi \me^{i \omega t} \, .
\end{equation}
Here we have used that, in TT gauge, proper time $\tau$ measured by an observer which is initially at rest is the same as coordinate time $t$, up to $\mathcal{O}\left(h^4\right)$~\cite{Maggiore:2007ulw}.
Therefore, we find for the observed electric field
\begin{equation}
	E_{\underline{2}} = \bF_{21} \delta u^1 = -\frac{i}{2} v_s A^{+} \bB \tan \left( \frac{\omega L}{2 v_s} \right) \me^{i \omega t} \, ,
\end{equation}
which is the \emph{exact} solution to first order in our perturbation scheme.
We illustrate this in \cref{fig:e_rod}, as a function of $\omega L$, and in units of $A^{+} \bB$.

\bigskip

We now aim to examine how the various mechanical approximations introduced in \cref{sec: mechanical deformations through gravitational waves} compare to this result.
The corresponding error estimates also have to be understood in units of $A^{+} \bB$.

For $\omega L\gg v_s$ the free-falling regime is approached.
As discussed above, this limit is obtained by going to the TT frame and neglecting the motion of the detector, i.e. setting $\delta x^\mathrm{TT}=0$.
In this case the observed electric field is given by $\delta F_{20}^{\rm TT}$ and one finds $E_{\underline{2}}=0$.
This approximation induces an error in the observable electric field that is of the order $\mathcal{O}(v_s)$. 
This is illustrated by the blue region in \cref{fig:e_rod}.
While clearly the mechanical resonances of the rod are not captured by the free-falling approximation, we find that it agrees with the exact result within the margin of error.

Similarly, for $\omega L \ll v_s$, we can take the rigid limit by setting $\delta x^{\mathrm{FN}} = 0$.
In this case, the only contribution to the observed electric field is due to the components $\delta F^{\mathrm{FN}}_{02}$ given in \cref{eq:F_FN_rod}.
This neglects terms of order $\mathcal{O}\left(\omega^3 L^3 / v_s^2\right)$ with respect to the exact result.
We illustrate the rigid limit in orange in \cref{fig:e_rod}.
The shaded region illustrates an improved estimate for the approximation error, $\mathcal{O}\left(\omega^3 L^3 / (\pi v_s)^2\right)$, where we take into account the first mechanical resonance at $\omega L = \pi v_s$.

Finally, let us comment on the long-wavelength limit, which is designed to approximate the regime $\omega L \ll 1$.
As we have pointed out in \cref{sec:coordinate_frames}, this is done by neglecting contributions to the metric in FN coordinates of orders beyond $\mathcal{O}(\omega^2 L^2)$, but taking into account the deformations of the detector at leading order in $\omega L$.
In this limit, following~\cite{Maggiore:2007ulw}, we find for the velocity of the observer
\begin{equation}
    \delta u^{\mathrm{FN}, 1} = - 2 i \omega L A^{+} \me^{i \omega t} \sum_{n=0}^{\infty} \frac{\omega^2}{\omega^2 - k_n^2 v_s^2} \frac{1}{k_n^2 L^2} \, ,
\end{equation}
where we have defined the resonant momenta $k_n = (2n + 1) \pi / L$.
Combining this result with the electric field~\eqref{eq:F_FN_rod} found in the FN frame, $E_{\underline{2}} = \delta F^{\mathrm{FN}}_{02}+\bF_{21} \delta u^{\mathrm{FN},1}$, we find the long-wavelength approximation of the observable electric field.
We note that this approximation neglects terms of order $\mathcal{O}(\omega^2 L^2)$ with respect to the exact result. 
This is shown in green in \cref{fig:e_rod}. 
Intriguingly, the long-wavelength approximation happens to coincide with the exact solution for a broad range of values for $\omega L$, even beyond its naive range of validity.
We suspect this to be an artefact of the highly symmetric configuration considered in this example, however.

\subsection{The observable magnetic field}

Our observations are further reinforced when considering the observable magnetic field.
The previous example highlights the relevance of carefully including the observer's four-velocity in the observable electric field.
Similarly, the spatial components of the observer's tetrad play a crucial role in the consistent treatment of the observable magnetic field pointing in the $x$-direction.
In TT gauge we find $\delta F_{23} = 0$, which in FN coordinates corresponds to
\begin{equation}
	\delta F^{\mathrm{FN}}_{23} = -\frac{i}{2} \omega x A^{+} \bB \me^{i \omega t} \, ,
\end{equation}
again leading to an apparent contradiction if both quantities are naively interpreted as the observed magnetic field.
Instead, in our coordinate-invariant framework the accurate expression for this field is, however,
\begin{equation}
	B_{\underline{1}} = \delta x^\lambda \left(\pd_\lambda \bF_{\mu \nu}\right) \overline{e}_{\underline{2}}^{\mu} \overline{e}_{\underline{3}}^{\nu}
					+ \delta F_{\mu \nu} \overline{e}_{\underline{2}}^{\mu} \overline{e}_{\underline{3}}^{\nu}
					+ \bF_{\mu \nu} \delta e_{\underline{2}}^{\mu} \overline{e}_{\underline{3}}^{\nu}
					+ \bF_{\mu \nu} \overline{e}_{\underline{2}}^{\mu}  \delta{e}_{\underline{3}}^{\nu} \, .
\label{eq:observed_B_field}
\end{equation}
Note that, here, the first term vanishes since the background field is static and homogeneous, $\pd_{\lambda} \bF_{\mu \nu} = 0$.
Therefore, in addition to the contribution of $\delta F_{\mu \nu}$, we need to determine the fluctuations of the tetrad, $\delta e_{\underline{2}}^{\mu}$ and $\delta e_{\underline{3}}^{\mu}$, whose dynamics are governed by \cref{eq:EOMTetrad}.
In our choice of perturbation scheme, we find
\begin{equation}
	\frac{\md}{\md t} \delta e^\mu_{\underline{a}} + \frac{1}{2}\eta^{\mu \rho}\left(\pd_\nu h_{\lambda\rho}+\pd_\lambda h_{\nu\rho}-\pd_\rho h_{\nu\lambda}\right) \overline{u}^\nu \overline{e}^\lambda_{\underline{a}} =  \delta a_\nu \overline{u}^\mu \overline{e}^\nu_{\underline{a}} \, ,
\end{equation}
where we have assumed that the sensor is freely rotating, $\omega^{\mu} = 0$.
We have also neglected terms involving the acceleration $\overline{a}^{\mu}$, given that the experiment is at rest in the unperturbed system, $\overline{a}^{\mu} = 0$.
Furthermore, it is sufficient to only solve the equations of motion for the spatial components of the tetrad, $\delta e_{\underline{a}}^{i}$, as there is no electric background field present, $\bF_{0i} = 0$.
In this case, the right hand side of the equations of motion vanishes, such that in TT gauge we obtain
\begin{equation}
	\delta e_{\underline{a}}^i = - \frac{1}{2} \tensor{{\htt}}{^i_j} \overline{e}_{\underline{a}}^j \, .
\end{equation}
This means that the fluctuations are either proportional to their background values, e.g.~$\delta e_{\underline{2}}^{i} \propto \overline{e}_{\underline{2}}^i$, or vanish, $\delta e_{\underline{3}}^{i} = 0$.
Therefore, we find that the observable magnetic field pointing in $x$-direction is vanishing too,
\begin{equation}
	B_{\underline{1}} = 0 \, .
\end{equation}
On the contrary, in FN coordinates we obtain a different motion of the tetrad,
\begin{equation}
	\delta e_{\underline{2}}^{i} = 0 \, , \enspace
	\delta e_{\underline{3}}^{i} = A^{+} \me^{i \omega t} \left( - \frac{i}{2} \omega x , 0 , \frac{1}{12} \omega^2 x^2 \right)^{i} \, .
\end{equation}
In this scenario, the contribution of the tetrad exactly cancels the nonvanishing field $\delta F^{\mathrm{FN}}_{23}$, such that the observable magnetic field again vanishes overall.
This alleviates the naive tension between both coordinate frames in a single unifying framework.

\section{The Spherical Cavity}
\label{sec: spherical cavity}

Let us now consider a more sophisticated example where a GW passes through a spherical electromagnetic cavity, i.e.~a hollow sphere in a homogeneous background magnetic field.
We begin with a planar GW of frequency $\omega$ propagating in the $z$-direction.
In the TT frame it is parametrised by
\begin{equation}
	\htt_{ij} = \frac{1}{2} \left[ h^{+} \begin{pmatrix}
							1 & i & 0 \\
							i & -1 & 0 \\
							0 & 0 & 0
							\end{pmatrix}
					+ h^{-} \begin{pmatrix}
							1 & -i & 0 \\
							-i & -1 & 0 \\
							0 & 0 & 0
							\end{pmatrix}
			\right] \me^{i \omega \left(t - z\right)} \, ,
\end{equation}
in terms of circular polarisation amplitudes $h^{\pm}$.
For simplicity, we will assume that the background magnetic field is small enough that the electromagnetic forces acting on the cavity wall can be neglected.
This allows us to determine the mechanical response of the cavity independently of the present electromagnetic field.
Overall, to find the electromagnetic fields that a suitable observer attached to the cavity would detect, we need to determine the mechanical deformations of the sphere and the excitations of the electromagnetic field modes due to the GW.
We will work out the exact answer in TT gauge, and later comment on the transformation into FN coordinates.

\subsection{Mechanical response}

Mechanical deformations of a spherical elastic body due to a GW have previously been studied in~\cite{Zhou:1995en, Lobo:1995sc, Coccia:1995yi, Coccia:1997gy}.
Along these lines, we have to solve the dynamics of elastic deformations given in \cref{eq:MechanicsEOM,eq:MechanicsBC}.
In the TT frame, they read
\begin{equation}
	\rho \pd_t^2 \delta \vec{x} = \left( \lambda + \mu \right) \nabla \left( \nabla \cdot \delta \vec{x} \right) + \mu \nabla^2 \delta \vec{x} \, ,
\label{eq:MechanicsEOMSphere}
\end{equation}
where the mechanical deformation, parametrised by $\delta \vec{x}$, is subject to the boundary condition
\begin{equation}
	\left. \left( \lambda \vec{e}_r \left( \nabla \cdot \delta \vec{x} \right) + 2 \mu \pd_r \delta \vec{x} + \mu \vec{e}_r \times \left(\nabla \times \delta \vec{x} \right) + \mu \vec{y} \right) \right\rvert_{r=R,R+\Delta R} = 0 \, .
\label{eq:MechanicsBCSphere}
\end{equation}
Here, $R$ denotes the inner radius and $\Delta R$ the wall thickness of the cavity, such that the boundary condition has to be satisfied at every point on both its inner and its outer wall.
Furthermore, the radial unit vector $\vec{e}_r$ is the normal vector on the sphere and we have introduced the source term $\vec{y}$ with components $y_i = h_{ij} e_{r,j}$.
The latter is proportional to the plane wave oscillations of the GW, $\vec{y} \propto \exp (i \omega t)$.
Therefore, the mechanical deformation $\delta \vec{x}$ must share this feature, such that we can make the ansatz~\cite{Maggiore:2007ulw}
\begin{equation}
	\delta \vec{x} = \left( \nabla \chi + i \nabla \times \vec{L} \phi + i \vec{L} \psi \right) \me^{i \omega t} \, .
\end{equation}
Here, $\vec{L}$ denotes the angular momentum operator, $\vec{L} = -i \vec{x} \times \nabla$.
In this parametrisation, the function $\chi$ characterises longitudinal waves, while both $\phi$ and $\psi$ characterise transverse waves.
Plugging this ansatz into the equations of motion~\eqref{eq:MechanicsEOMSphere}, we obtain
\begin{equation}
	\left( \nabla^2 + p^2 \right) \chi = 0 \, , \enspace
	\left( \nabla^2 + q^2 \right) \phi = 0 \, , \enspace
	\left( \nabla^2 + q^2 \right) \psi = 0 \, , \enspace
\end{equation}
where we have defined
\begin{equation}
	p^2 = \frac{\rho}{\lambda + 2\mu} \omega^2 \, , \enspace q^2 = \frac{\rho}{\mu} \omega^2 \, .
\label{eq:P_Q}
\end{equation}
Due to the spherical symmetry of the problem, the general solution for each contribution can then be written as
\begin{equation}
	\chi (r, \theta, \varphi) = \sum_{l=0}^{\infty} \sum_{m=-l}^l \left( \chi_{lm} j_l (pr) + \tilde{\chi}_{lm} y_l (pr) \right) Y_l^m (\theta, \varphi) \, ,
\label{eq:ChiSphericalExpansion}
\end{equation}
and similarly for $\phi$ and $\psi$, where $p$ is replaced by $q$.
Here, $Y_l^m$ denote the spherical harmonics and $j_l$ and $y_l$ are the spherical Bessel functions of the first and second kind, respectively.
In practice, the mode functions $\chi_{lm}$, $\tilde{\chi}_{lm}$, \textit{etc.}~are determined by the boundary condition~\eqref{eq:MechanicsBCSphere}.
These mode functions completely determine the mechanical deformation $\delta \vec{x}$, and in particular the mechanical resonances, of the spherical cavity.
We present a detailed computation of $\delta \vec{x}$ in \cref{app:DeformationCoefficients}.

In summary, we find that the GW only couples to modes with $m = 2$, i.e.~any other mode function vanishes, such that the expansions in spherical harmonics~\eqref{eq:ChiSphericalExpansion} for $\chi$, $\phi$ and $\psi$ only start at the second order, $l = 2$. 
In explicit calculations, the expansion has to be truncated at some finite order $l \geq 2$.
As we will see in the following section, the resonant mechanical excitations occur at lower frequencies compared to the electromagnetic resonances.

\subsection{Electromagnetic response}
\label{sec:ElectromagneticExcitations}

Let us now determine the electromagnetic fields inside the spherical cavity that may be excited by the incoming GW in the vicinity of the homogeneous magnetic background field, $\bvB$.
It is convenient to parametrise this background field as
\begin{equation}
	\bvB = \bB^{-} \vec{e}^{-} + \bB^0 \vec{e}^0 + \bB^{+} \vec{e}^{+} \, ,
\end{equation}
where we have defined the orthonormal spherical basis
\begin{equation}
	\vec{e}^{\pm} = \mp \frac{1}{\sqrt{2}} \left( \vec{e}_x \pm i  \vec{e}_y \right) \, , \enspace
	\vec{e}^0 = \vec{e}_z \, .
\end{equation}
In this basis, the spatial components of the effective current~\eqref{eq:jeff} read
\begin{equation}
	\vec{j}_{\mathrm{eff}} = -\omega \left( \bB^{-} h^{+} \vec{e}^{+} - \bB^{+} h^{-} \vec{e}^{-} \right) \me^{i \omega \left(t - z\right)} \, .
\end{equation}
This current acts as the source for the electromagnetic field perturbations governed by the inhomogeneous Maxwell's equations at linear order~\eqref{eq:EOMFirstOrder}.
In practice, to find the electromagnetic field excitations from \cref{eq:EOMFirstOrder}, we can split these into a bulk and a boundary contribution, $\delta F_{i0} = E^{\mathrm{blk}}_i + E^{\mathrm{bnd}}_i$ and $\delta F_{ij} = \epsilon_{ijk} \left(B^{\mathrm{blk}}_k + B^{\mathrm{bnd}}_k\right)$, respectively.
In this decomposition, the bulk field is an arbitrary solution to the inhomogeneous equation involving the effective current $\vec{j}_{\mathrm{eff}}$.
The boundary field solves the homogeneous equations while guaranteeing that the boundary conditions are met.
In our scenario, one possible solution for the bulk fields is given by
\begin{align}
	\vec{E}_{\mathrm{blk}} &= - \frac{z}{2} \vec{j}_{\mathrm{eff}} = \frac{\omega z}{2} \left(\bB^{-} h^{+} \vec{e}^{+} - \bB^{+} h^{-} \vec{e}^{-} \right) \me^{i \omega \left(t - z\right)} \, , \\
	\vec{B}_{\mathrm{blk}} &=  \frac{1 - i \omega z}{2} \left(\bB^{-} h^{+} \vec{e}^{+} + \bB^{+} h^{-} \vec{e}^{-} \right) \me^{i \omega \left(t - z\right)} \, .
\end{align}
Again, by construction, these fields do not satisfy the boundary condition~\eqref{eq:BCCoordinateIndependent}.
This is instead guaranteed by the boundary field contribution, which can be written as (see, e.g., \cite{Olaussen:1981pc})
\begin{align}
	\vec{E}_{\mathrm{bnd}} &= i \nabla \times \vec{L} \zeta + i \vec{L} \eta \, , \label{eq:Ebnd}\\
	\vec{B}_{\mathrm{bnd}} &= - \omega \vec{L} \zeta - \frac{1}{\omega} \nabla \times \vec{L} \eta \, , \label{eq:Bbnd}
\end{align}
where the functions $\zeta$ and $\eta$ both satisfy
\begin{equation}
	\left( \nabla^2 + \omega^2 \right) \zeta = 0 \, , \enspace
	\left( \nabla^2 + \omega^2 \right) \eta = 0 \, .
\end{equation}
Similar to the mode functions of the mechanical deformations in \cref{eq:ChiSphericalExpansion}, both functions can again be expanded in terms of spherical harmonics,
\begin{equation}
	\zeta(r, \theta, \varphi) = \sum_{l=0}^{\infty} \sum_{m=-l}^{l} \zeta_{lm} j_l (\omega r) Y_l^m (\theta, \varphi) \, ,
\end{equation}
and similarly for $\eta$.
However, there are no contributions involving spherical Bessel functions of the second kind, $y_l(\omega r)$, because the field needs to be regular at the origin.
Similar to the mechanical deformations, the mode functions $\zeta_{lm}$ and $\eta_{lm}$ are determined by the appropriate boundary conditions that the electromagnetic fields have to satisfy.
In our scenario, a fully covariant form of the boundary condition is given in \cref{eq:BCCoordinateIndependent}, which in our perturbation scheme reads
\begin{equation}
	\left. \left( \delta x^{\lambda} \left(\pd_{\lambda} \bF_{\mu \nu}\right) \overline{e}_{\underline{1},\underline{2}}^{\mu} \overline{u}^{\nu}
	+ \delta F_{\mu \nu} \overline{e}^{\mu}_{\underline{1},\underline{2}} \overline{u}^{\nu}
	+ \bF_{\mu \nu} \delta e^{\mu}_{\underline{1},\underline{2}} \overline{u}^{\nu}
	+ \bF_{\mu \nu} \overline{e}_{\underline{1},\underline{2}}^{\mu} \delta u^{\nu}
	\right) \right\rvert_{r=R} = 0 \, .
\end{equation}
Here, the vectors $e_{\underline{1}}^{\mu}$ and $e_{\underline{2}}^{\mu}$ are tangential to the surface of the spherical cavity, and $\bF_{\mu \nu}$ encodes the static and homogeneous magnetic background field, such that $\pd_{\lambda} \bF_{\mu \nu} = 0$ and $\bF_{\mu \nu} \overline{u}^{\nu} = 0$.
Therefore, the boundary condition simplifies to
\begin{equation}
	\left. \left( \vec{E}_{\mathrm{bnd}} + \vec{E}_{\mathrm{blk}} + i \omega \delta \vec{x} \times \bvB \right) \right\rvert_{\parallel, r=R} = 0 \, ,
\label{eq:ElectromagneticBCSphericalCavity}
\end{equation}
where the subscript $\parallel$ indicates that only the components tangential to the unperturbed surface of the sphere are to be considered.
The mode functions $\zeta_{lm}$ and $\eta_{lm}$ enter this boundary condition through the contribution of $\vec{E}_{\mathrm{bnd}}$ and are hence entirely determined by the remaining terms.
We present the details of this computation in \cref{app:ElectromagneticCoefficients}.
In practice, we truncate the expansion in spherical harmonics at some finite order $l$.

In summary, the electromagnetic field inside the cavity that is excited by the incoming GW is given by a bulk and a boundary contribution.
The latter explicitly couples the electromagnetic to the mechanical response of the cavity, entering the electromagnetic fields that can be measured by an external observer.

\subsection{The observable magnetic field}

The electromagnetic excitations of a cavity are typically read out using antennas and pickup loops.
In particular, magnetic fields are measured using a pickup loop, which is essentially a conducting wire that encloses a certain area $A$.
Ideally, it is oriented such that an excitation of the cavity leads to a varying magnetic flux through this area, which, in turn, induces an electric current in the wire that can be measured.

In this work, for simplicity, we assume a rigid infinitesimal pickup loop with a magnetic flux
\begin{equation}
	\Phi = A F_{\mu \nu} e^{\mu}_{\underline{1}} e^{\nu}_{\underline{2}} \, ,
\end{equation}
where $e^{\mu}_{\underline{1}}$ and $e^{\nu}_{\underline{2}}$ are spatial components of the tetrad that is following the worldline of the loop.
Geometrically, they span the plane that the pickup loop area is embedded in.
In this simple setup, we identify the observable magnetic field as
\begin{equation}
	B_{\mathrm{obs}} = F_{\mu \nu} e^{\mu}_{\underline{1}} e^{\nu}_{\underline{2}} \, .
\end{equation}
While this is a fully covariant form of the observable magnetic field through the pickup loop, in our perturbation scheme it reads
\begin{equation}
	B_{\mathrm{obs}} = \delta x^{\lambda} \left(\pd_{\lambda} \bF_{\mu \nu}\right) \overline{e}^{\mu}_{\underline{1}} \overline{e}^{\nu}_{\underline{2}}
	+ \delta F_{\mu \nu} \overline{e}^{\mu}_{\underline{1}} \overline{e}^{\nu}_{\underline{2}}
	+ \bF_{\mu \nu} \delta e^{\mu}_{\underline{1}} \overline{e}^{\nu}_{\underline{2}}
	+ \bF_{\mu \nu} \overline{e}^{\mu}_{\underline{1}} \delta e^{\nu}_{\underline{2}} \, .
\end{equation}
If we again use that the background field is static and homogeneous, $\pd_\lambda \bF_{\mu \nu} = 0$, we can write the observable magnetic field as
\begin{equation}
	B_{\mathrm{obs}} =
	\overline{\vec{e}}_{\underline{1}} \times \overline{\vec{e}}_{\underline{2}} \cdot \left( \vec{B}_{\mathrm{blk}} + \vec{B}_{\mathrm{bnd}} \right)
	+ \left( \delta \vec{e}_{\underline{1}} \times \overline{\vec{e}}_{\underline{2}} + \overline{\vec{e}}_{\underline{1}} \times \delta \vec{e}_{\underline{2}} \right) \cdot \bvB \, .
\label{eq:BobsExplicit}
\end{equation}
This is the general coordinate-independent expression for the magnetic field that is read out by the pickup loop.
It is attached to the cavity such that its four-velocity component, $e_{\underline{0}}^{\mu} = u^{\mu}$, coincides with the wall's at the attachment point.
If, in addition, the pickup loop is only loosely attached to the cavity, the wall will not exert any torque.
Let us therefore, for simplicity, assume that the pickup loop is attached at its center of mass, such that the overall torque vanishes.\footnote{In principle, corrections to this assumption can be incorporated by considering the mass distribution and a specific attachment point of any given pickup loop. One could also consider a scenario where the loop is rigidly attached to the cavity wall, such that the motion of the observer's tetrad is entirely fixed by the mechanical deformations of the cavity.}
In this scenario, the equations of motion of the observer's tetrad are given by \cref{eq:EOMTetrad} with vanishing rotation, $\omega^{\mu} = 0$.
If we only focus on the spatial-spatial components of this tetrad, $\delta e_{\underline{a}}^{i}$, the terms proportional to the acceleration vanish, since $\overline{u}_{\mu} \overline{e}_{\underline{a}}^{\mu} = 0$.
The tetrad's equations of motion in this case simplify to
\begin{equation}
	\frac{\md}{\md t} \delta e_{\underline{a}}^{i} + \frac{1}{2} \pd_t \tensor{h}{^i _j} \overline{e}_{\underline{a}}^{j} = 0 \, ,
\end{equation}
such that we obtain
\begin{equation}
	\delta e_{\underline{a}}^{i} = - \frac{1}{2} \tensor{h}{^i _j} \overline{e}_{\underline{a}}^{j} \, .
\end{equation}
We therefore find that tetrad contribution to the observable magnetic field is proportional to $\left( \delta \vec{e}_{\underline{1}} \times \overline{\vec{e}}_{\underline{2}} + \overline{\vec{e}}_{\underline{1}} \times \delta \vec{e}_{\underline{2}} \right)_{i} = \frac{1}{2} h_{ij} \overline{e}_{\underline{3}}^{j} \,$, where we have used that $\overline{\vec{e}}_{\underline{1}} \times \overline{\vec{e}}_{\underline{2}} = \overline{\vec{e}}_{\underline{3}}$.
Overall, if we denote by $\vec{d} = \overline{\vec{e}}_{\underline{3}}$ the direction in which the loop reads out the magnetic field, we obtain
\begin{equation}
	\left( \delta \vec{e}_{\underline{1}} \times \overline{\vec{e}}_{\underline{2}} + \overline{\vec{e}}_{\underline{1}} \times \delta \vec{e}_{\underline{2}} \right) \cdot \bvB
	= \frac{1}{2} \left(\bB^{+} d^{+} h^{-} + \bB^{-} d^{-} h^{+} \right) \me^{i \omega \left(t - z\right)} \, ,
\end{equation}
as the second term entering the observable magnetic field. 
This result is exact to linear order in the metric perturbation, and we will illustrate the magnetic field measured by an external observer in the following.

\subsection{Results}

\begin{figure}[t]
	\centering
	\includegraphics[width=0.7\textwidth]{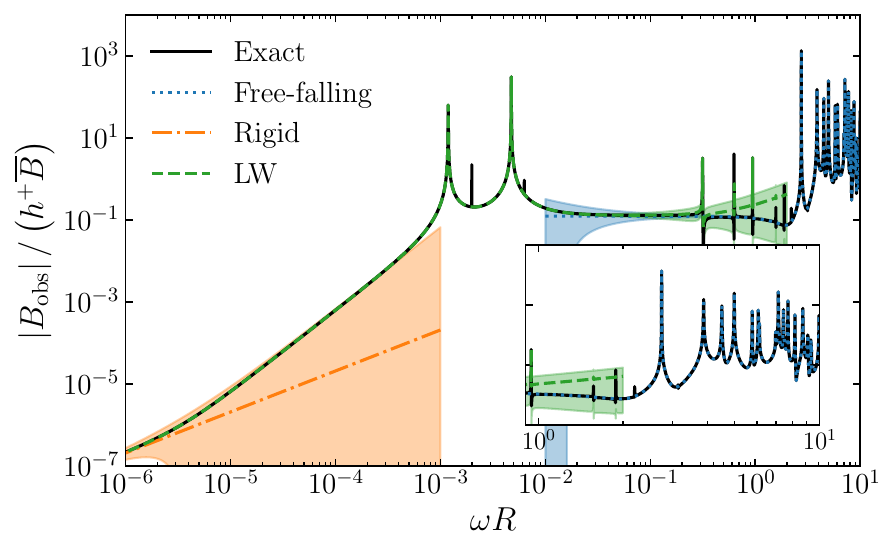}
	\caption{Amplitude of the observed magnetic field measured by a pickup loop attached to the spherical cavity of radius $R$, as a function of $\omega R$.
    The inset illustrates the electromagnetic resonances at large $\omega R$.
    Here, we include terms up to $l=5$ in the expansion in spherical harmonics.
    The sound velocity is chosen to be $v_s = 10^{-3}$, following from a ratio of Lam\'e parameters of $\lambda / \rho = 2\mu / \rho = 2 \times 10^{-6}$.
    Furthermore, the relative wall thickness is $\Delta R / R = 1~\%$, and we have chosen a plus polarisation of the GW, $h^{+} = 1$ and $h^{-} = 0$.
    The incoming GW is propagating in the $z$-direction and the magnetic background field is $\bvB = -1/2 \vec{e}_x + \sqrt{3}/2 \vec{e}_z$.
    The pickup loop is then located at angles of $\theta = 7 \pi / 12$ and $\varphi = 0$, while the sensor points to the $y$-direction, $\vec{d} = \vec{e}_y$.
    We note that the shown error estimates include numerical factors, reading $\mathcal{O}(2 \, v_s / (\omega R))$ (blue), $\mathcal{O}\left(\frac{1}{15} \, \omega^2 R^2 / v_s^2\right)$ (orange), and $\mathcal{O}\left(\frac{1}{5} \, \omega R\right)$ (green), respectively.}
\label{fig:b_cavity_thinwall}
\end{figure}

\begin{figure}[t]
	\centering
	\includegraphics[width=0.7\textwidth]{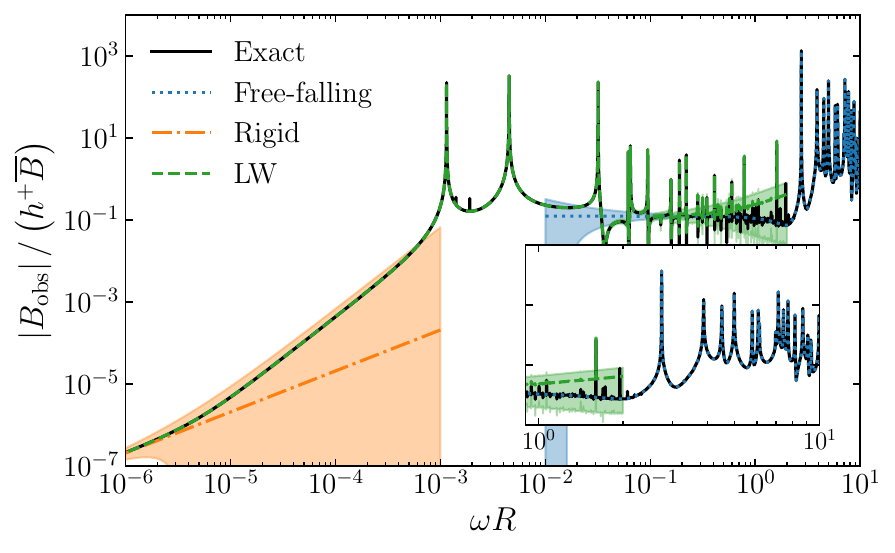}
	\caption{Amplitude of the observed magnetic field measured by a pickup loop attached to the spherical cavity of radius $R$, as a function of $\omega R$.
	The relative wall thickness is chosen as $\Delta R / R = 10~\%$, while all other parameters are the same as in \cref{fig:b_cavity_thinwall}.
	The inset illustrates the electromagnetic resonances at large $\omega R$.}
\label{fig:b_cavity_thickwall}
\end{figure}

At this point, we have determined all contributions to the observable magnetic field~\eqref{eq:BobsExplicit}.
This result is exact to linear order in the perturbation scheme, and we illustrate two example configurations of the magnetic field strength as a function of $\omega R$ in \cref{fig:b_cavity_thinwall,fig:b_cavity_thickwall}, in units of $h^{+} \bB$.
These configurations differ in their choice of the cavity's thickness, but are otherwise generic in the sense that no contribution to the magnetic field is parametrically enhanced or suppressed, e.g.~due to an enhanced symmetry.
Furthermore, while the above computation of the observable magnetic field is performed in TT gauge, we give an overview of the analogous calculation in FN coordinates in \cref{app:SphericalCavityFN}.

Similar to the thin rod, it is useful to consider the signal in different frequency regimes together with the appropriate approximations and their corresponding error estimates, again in units of $h^{+} \bB$.
At small frequencies, for $\omega R \ll 1$, FN coordinates are a natural choice to employ the long-wavelength approximation laid out in \cref{sec:coordinate_frames}.
In particular, in the regime $\omega R\lesssim v_s^2 = 10^{-6}$, the signal approaches the rigid approximation where the mechanical response of the detector is neglected, $\delta \vec{x}^{\mathrm{FN}} = 0$.\footnote{We note that, for the exact result in TT gauge to reproduce the correct behaviour in the rigid limit, we have to include terms of the mechanical detector response, $\delta \vec{x}^{\mathrm{TT}}$, up to $l=3$.}
This approximation parametrically behaves as $\mathcal{O}(\omega R)$, and is shown in orange in \cref{fig:b_cavity_thinwall,fig:b_cavity_thickwall}.
It turns out that in this example the signal is merely due to the perceived rotation of the observer's tetrad, while the effective current only contributes as $\mathcal{O} \left(\omega^2 R^2\right)$.
The dominant error is of the order $\mathcal{O} \left(\omega^2 R^2/v_s^2\right)$.
This is due to neglecting the detector's motion, and is illustrated by the orange-shaded region.

At larger frequencies, clearly, the motion of the detector has to be taken into account.
In the regime $v_s \lesssim \omega R \ll 1$, we still expect FN coordinates to be a suitable choice of frame, and similarly the long-wavelength approximation to apply.
Therefore, we include the leading-order terms in $\omega R$ of the detector's motion (see also \cref{sec: mechanical deformations through gravitational waves}).
This motion is described by the spheroidal modes with angular momentum $l=2$~\cite{Maggiore:2007ulw}, which, roughly speaking, correspond to deformations of the sphere into an ellipsoid~\cite{Mazurenko_2007}.
As can be seen in \cref{fig:b_cavity_thinwall,fig:b_cavity_thickwall}, where the long-wavelength limit is shown in green, the result in FN coordinates is in good agreement with the exact one obtained in TT gauge, up to frequencies comparable to the inverse size of the cavity, $\omega R \sim 1$.
In the regime where the long-wavelength approximation breaks down, their difference behaves as $\mathcal{O} (\omega R)$ and is caused by neglecting subleading terms contributing to the cavity's motion.
In fact, these approximation errors already appear within the range of validity of the long-wavelength limit, i.e.~at frequencies of $\omega R \sim 10^{-3}-10^{-2}$.
Here, the exact result features two subleading mechanical resonances (shown in black).
Our parametric error estimates do not account for such resonances, but we remark that their suppression is visible in the sense that they appear very thin compared to the $l=2$ spheroidal resonances.
In \cref{fig:b_cavity_thinwall,fig:b_cavity_thickwall} we have included terms up to $l=5$, both in the mechanical and electromagnetic expansion in spherical harmonics.

Furthermore, we observe that the relative importance of the $l=2$ spheroidal resonances beyond the first two is suppressed.
These include excitations along the radial direction, which is why they are separated from the first two by the inverse of the relative wall thickness.
This can be seen by comparing \cref{fig:b_cavity_thinwall,fig:b_cavity_thickwall}, where we show the same configuration for a relative wall thickness of $\Delta R / R = 1~\%$ and $\Delta R / R = 10~\%$, respectively.
As these excitations become more suppressed, the signal quickly approaches the free-falling limit.
A similar behaviour has been observed in~\cite{Berlin:2023grv}.

Finally, at even higher frequencies, $\omega R \gg v_s$, we expect the free-falling approximation to be valid, which is obtained by neglecting the mechanical detector response in the TT frame, $\delta \vec{x}^{\mathrm{TT}} = 0$.
In \cref{fig:b_cavity_thinwall,fig:b_cavity_thickwall}, this approximation is shown in blue.
It neglects terms which are of the order $\mathcal{O}\left(v_s / (\omega R)\right)$, and, for $\omega R \lesssim 1$, it is dominated by the tetrad's motion while the contribution arising from the effective current is of the order $\mathcal{O} (\omega R)$.
Beyond that, at $\omega R\gtrsim 1$ the resonant electromagnetic excitations appear.
In this regime the free-falling approximation is in excellent agreement with the exact result.

We close this discussion by noting that, in all regimes, the approximation errors are enhanced (or suppressed) by a factor $\sim 1 / (\omega R)$ compared to the previous example of the thin rod.
For instance, the dominant approximation error caused by the rigid approximation is now of the order $\mathcal{O}\left(\omega^2 R^2 / v_s^2\right)$ instead of $\mathcal{O} \left( \omega^3 L^3 / v_s^2 \right)$ found in \cref{sec: thin rod}.
This is due to the electric field contribution that is caused by the apparent motion of the detector, i.e.~the third term of \cref{eq:ElectromagneticBCSphericalCavity}, which has to be partially compensated by the boundary field proportional to $\eta$.
The latter, in turn, leads to the enhancement (or suppression) of the corresponding magnetic field (cf.~\cref{eq:Ebnd,eq:Bbnd}).
In fact, it is sufficient to only include the $\eta$ mode with $l=1$ in the long-wavelength approximation, with other modes contributing as $\mathcal{O} (\omega R)$ or higher.

\section{Conclusions}
\label{sec: conclusions}

In this work, we have presented a coordinate-invariant framework to consistently describe the signal that a GW leaves in an experiment where it is converted into electromagnetic radiation via the inverse Gertsenshtein effect.
More precisely, we have given manifestly coordinate-invariant expressions for the observed electric and magnetic fields that an external observer would measure in this scenario.
After choosing a perturbation scheme, we have derived the equations of motion that govern the fluctuations of the electromagnetic fields around their background values upon the arrival of a GW.
These correlate the electromagnetic fluctuations and the perceived motion of the detector in any given coordinate frame.
We have further defined a free-falling and a rigid approximation where this motion is suppressed either in TT gauge or FN coordinates, in the regimes $\omega L \gg v_s$ and $\omega L \ll v_s$, respectively.
We have illustrated our framework in two examples, a thin rod and a spherical cavity, to demonstrate that our formalism indeed leads to coordinate-independent results.
However, the discussion of the spherical cavity highlights that carefully working out the formalism for a given scenario may require a substantial amount of computational effort, because the mechanical and electromagnetic properties as well as the sensors of the experiment have to be carefully characterised.
We expect that the description of more complicated detector setups require an implementation of the approximations mentioned before.
Along these lines, we want to conclude this discussion by outlining some practical approaches for a range of experimental scenarios.

The free-falling limit is the natural choice for experimental setups where the GW frequency and the characteristic length scale of the experiment are comparable, $\omega \sim 1 / L$.
This typically covers electromagnetic cavities that rely either on the conversion of GWs to a resonant excitation in an external magnetic field, or on the conversion between different resonances.
Compared to the exact result, the error of this approximation is suppressed by the sound velocity inside the detector material, which typically is of the order $v_s \sim \mathcal{O}(10^{-5})$.
These experiments mostly focus on narrowband searches, where a large quality factor $Q$ of the resonant excitations (at frequencies $\omega_{\mathrm{res}}$) is leveraged to increase their sensitivity.
In this case, one is perhaps only interested in the signal region close to the resonant frequencies.
Here, the electromagnetic fluctuations $\delta F_{\mu \nu}$ are parametrically enhanced with respect to the terms including the measurement details, such as the location of the sensor, $\delta x$, and its direction, $\delta e_{\underline{a}}$.
For a realistic antenna we would not expect this to be drastically different, either.
In this case, one may approximate the observed fields by the fluctuations $\delta F_{\mu \nu}$, while the approximation errors are parametrically suppressed by a factor of $\sim 1/Q$ or $\sim \lvert \omega_{\mathrm{res}} - \omega \rvert / \omega_{\mathrm{res}} \,$.
In practice, the computation of the signal then reduces to determining the overlap function between the effective current and the electromagnetic resonant modes of the experiment (see, e.g.,~\cite{Berlin:2021txa} for details).
We remark that this estimate, strictly speaking, is only applicable in TT gauge, as in other frames also the overlap with the unsuppressed detector motion has to be taken into account.

On the contrary, sensitivity estimates for experimental setups operating at relatively low frequencies, $\omega \ll 1/L$, typically feature more intricacies.
These rely on, e.g., lumped circuits or quasi-degenerate modes with $\Delta \omega_{\mathrm{res}} \ll 1/L$, and support both narrow- and broadband operation modes.
That said, they typically are sensitive to the transition region between the rigid and the free-falling limit, as $\omega L \sim v_s \,$.
As a silver lining, it is safe to say that these experiments allow for a description in the long-wavelength approximation.
However, so far this approximation has only been employed for the MAGO experiment~\cite{Ballantini:2003nt,Berlin:2023grv}, while estimates for lumped-circuit experiments are based on a fully rigid approximation~\cite{Domcke:2022rgu,Domcke:2023bat}.
Although the rigid and freely-falling limit only approximate the exact GW signal, they nevertheless provide a valuable crosscheck since any (somewhat more rigorous) treatment of the detector's mechanical response must converge to the corresponding limit for $\omega L \ll v_s$ and $\omega L \gg v_s$, respectively.

In future work, it would be interesting to examine how the analysis of mechanical resonances presented here and in~\cite{Ballantini:2003nt,Berlin:2023grv} may be generalised.
So far, these estimates make strong assumptions about the mechanical behaviour of the antenna, which in our example of the spherical cavity is treated as an infinitesimal (and hence rigid) freely-rotating pickup loop.
These assumptions, albeit somewhat unrealistic, significantly simplify the analysis of the detector's response, because its motion does not enter the observable electromagnetic fields immediately but only as a contribution to the boundary conditions.
In this case, at least close to a mechanical resonance, a description via overlap functions between the incoming GW and the coupling of the resonance to the electromagnetic modes appears suitable (see also~\cite{Berlin:2023grv}).
On the other hand, it seems reasonable that an analogous decoupling behaviour occurs in the opposite regime involving an extended feeble antenna that is freely falling.
Along these lines, one could perhaps investigate sensitivity improvements through deliberately designed mechanical setups of measurement antennas optimised for GW detection.

\section*{Acknowledgements}

We thank Valerie Domcke, Sebastian A.R.~Ellis and Joachim Kopp for discussions initiating this project, and Kristof Schmieden and Matthias Schott for collaboration on related topics.
We further thank the participants of the UHF-GW workshop \textit{Where to next?} for ample feedback on our work as well as the CERN theory department for their hospitality.
SS and PS are supported by the Cluster of Excellence \textit{Precision Physics, Fundamental Interactions and Structure of Matter} (PRISMA$^+$ EXC 2118/1) funded by the German Research Foundation (DFG) within the German Excellence Strategy (Project No.~390831469).

\appendix

\section{The Elastic Conductor}
\label{app:ElasticConductor}

Here, we aim to give a brief discussion of an ideal conductor in the vicinity of a gravitational field.
We assume that the conductor represents a solid body that is described by the Carter-Quintana approach~\cite{Carter:1972} (see also \cite{Hudelist:2022ixo,Beig:2023pka}).
Mathematically, the dynamics of the elastic body are characterised by a so-called deformation map, $f:\mathcal{M} \to \mathcal{B}$, from the spacetime manifold $\mathcal{M}$ into the body's three-dimensional manifold $\mathcal{B}$.
The configuration is such that there exists a timelike unit vector field, $u$, the conductor's four-velocity, with $\md f(u) = 0$.

Inside an ideal conductor, the electric field measured by any observer moving along has to vanish.
This requirement can be written as $F(u, \cdot)=0$, where $F$ is the electromagnetic 2-form.
Clearly, $F$ needs to satisfy the homogeneous Maxwell's equations, $\md F = 0$, where $\md$ now denotes the exterior derivative.
In fact, we can obtain $F$ for any 2-form defined on the body's manifold, $F^{\mathcal{B}}$, that is exact, $\md F^{\mathcal{B}}=0$.
This solution is given by the pullback by $f$, i.e.~$F = f^{*} F^{\mathcal{B}}$.\footnote{This is because $F(u,\cdot)=F^\mathcal{B}(\md f(u),\md f(\cdot))=F^\mathcal{B}(0,\md f(\cdot))=0$ and the exterior derivative commutes with the pullback.}
Conversely, for every $F$ there exists such an $F^{\mathcal{B}}$~\cite{Beig:2023pka}.

In this way, we have recovered the concept of ``flux-freezing" in an elastic conductor, a phenomenon well known in ideal magnetohydrodynamics~\cite{Alfven:1943}.
It describes the effect that in a fluid offering no electrical resistance there is no relative motion between the magnetic field lines and the material.
In our approach, this constant magnetic field configuration is characterised by $F^{\mathcal{B}}$.
As we will show below, the magnetic flux through any patch on the conductor's surface is constant in time.
This is a generalisation of $\vec{B}_{\bot} = \mathrm{const}$.
This is known as ``flux-conservation" in magnetohydrodynamics, and is a direct consequence of flux-freezing.

\subsection*{Boundary Conditions}

\begin{figure}[t]
    \centering
    \includegraphics[width=.4\textwidth]{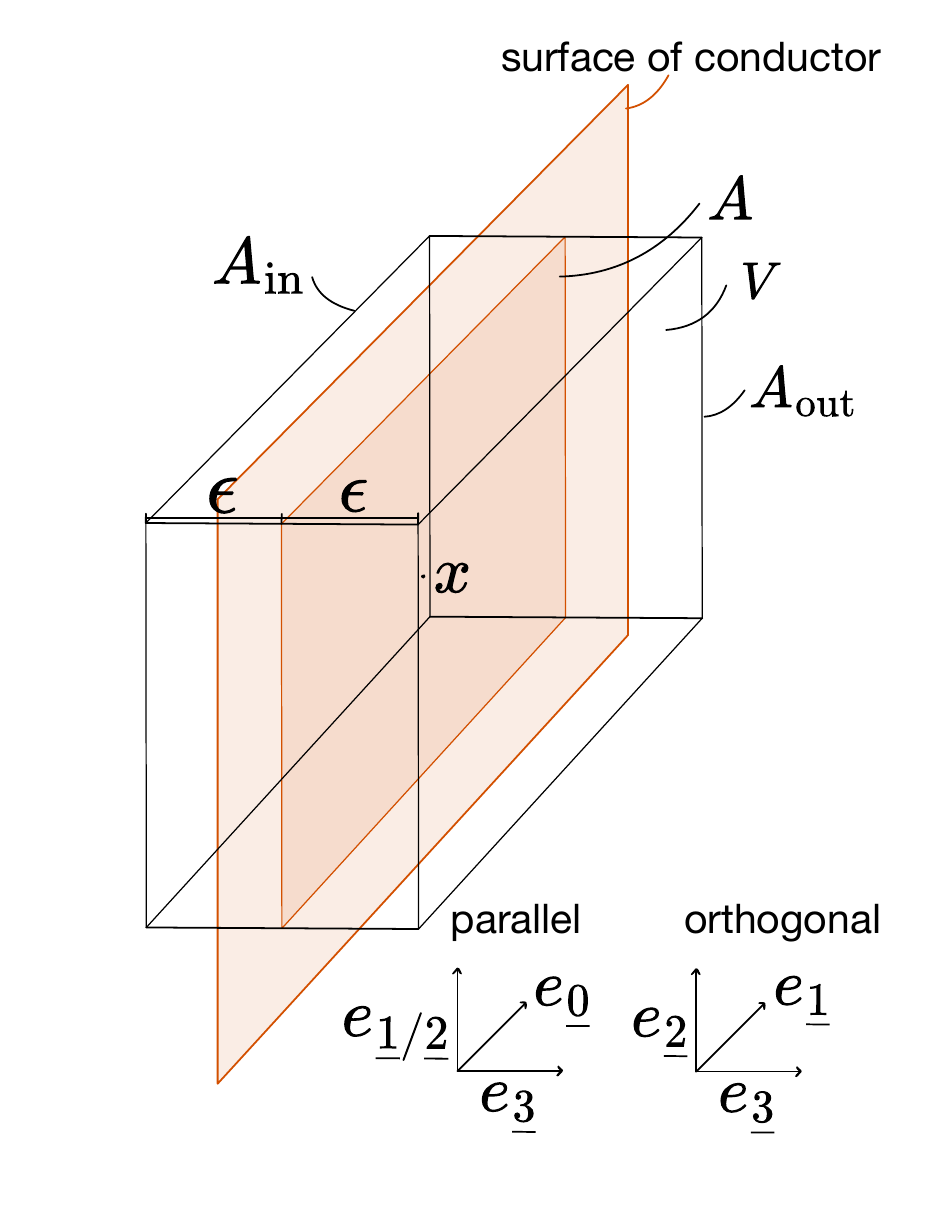}
    \caption{Sketch of the integration contours in  a point $x$ on the conductor's surface. The coordinate systems indicate the directions tangential to the 3-volume in $x$ used when considering field components parallel and orthogonal to the surface.}
    \label{fig: boundary condition} 
\end{figure}

We now want to use this setup to give a brief derivation of the boundary conditions that electromagnetic fields have to satisfy at the conductor's surface, i.e.~at the interface between the conducting material and vacuum.
A more detailed (and more rigorous) proof for the general case involving two arbitrary media can be found in~\cite{Rawson-Harris:1972nfp}.\footnote{Indeed, Ref.~\cite{Rawson-Harris:1972nfp} claims that this proof should only be valid for time-independent solutions, while referring to a set of equations from~\cite{Synge:1960ueh}. However, we think of these as Stokes' theorem applied to infinitesimal regions of spacetime, such that this restriction of the proof is not necessary.}

We begin by considering a spacetime point $x$ on the conductor's surface and a positively-oriented tetrad $e_{\underline{\alpha}}$, such that $e_{\underline{0}}=u$ is the conductor's four-velocity, $e_{\underline{1}}$ and $e_{\underline{2}}$ are tangential to the conductor's surface and $e_{\underline{3}}$ is normal to the surface, pointing away from the conductor.
We further define a three-dimensional volume $V$ that includes $x$ and, importantly, is tangential to $e_{\underline{0}}$, $e_{\underline{3}}$ and either $e_{\underline{1}}$ or $e_{\underline{2}}$ at the point $x$.
The boundary part of this volume that lies outside (inside) the conductor we denote by $A_{\mathrm{out}}$ ($A_{\mathrm{in}}$), such that the entire boundary is the union of both contributions, $\partial V = A_{\mathrm{out}} \cup A_{\mathrm{in}}$.
The intersection of $V$ with the conductor's surface is further denoted by $A$.
We will now study two limiting procedures that, if both are combined, shrink the volume $V$ into the point $x$.
The first limit takes the extent of $V$ in the direction of $e_{\underline{3}}$ to zero, schematically denoted by $\epsilon \to 0$.
This indeed implies $V \to 0$, as well as $A_{\mathrm{out}} \to A$ and $A_{\mathrm{in}} \to A$.
The second limit finally takes $A \to 0$.
We illustrate this scenario in \cref{fig: boundary condition}.

Now recall that the homogeneous Maxwell's equations written in terms of the 2-form $F$ read
\begin{equation}
	\md F = 0 \, .
\end{equation}
Therefore, using Stokes' theorem, we can write
\begin{equation}
    0 = \int_V \md F = \oint_{\partial V} F = \int_{A_\mathrm{out}} F + \int_{A_\mathrm{in}} F \, ,
\end{equation}
where we have assumed that both surfaces $A_{\mathrm{out}}$ and $A_{\mathrm{in}}$ are oriented the same way as the total boundary $\partial V$.
In the limit $\epsilon \to 0$, either $A_{\mathrm{out}}$ or $A_{\mathrm{in}}$ is then oriented according to $A$, while the other is pointing in the opposite direction.
Without loss of generality, we take the orientation of $A_{\mathrm{out}}$ to be equal to $A$ in this regime, such that we obtain the condition
\begin{equation}
	\int_A F^{\mathrm{out}} - \int_A F^{\mathrm{in}} = 0 \, ,
\end{equation}
where the superscripts indicate how the limit in $F$ is to be taken as $\epsilon \to 0$.
Indeed, for infinitesimally small $A$, the tetrad we have defined in $x$ can be extended to all points in $A$, in which case the above integrals can be rewritten as
\begin{equation}
	\int_A \md A \, F^{\mathrm{out}}_{\mu \nu} e^{\mu}_{\underline{0}} e^{\nu}_{\underline{1}, \underline{2}}
	- \int_A \md A \, F^{\mathrm{in}}_{\mu \nu} e^{\mu}_{\underline{0}} e^{\nu}_{\underline{1}, \underline{2}}
	= 0 \, ,
\end{equation}
where $\md A$ is the surface measure that is induced by the spacetime metric.\footnote{In Ref.~\cite{Rawson-Harris:1972nfp} this integrand is also written as $F^{\mathrm{out}}_{\mu \nu} e^{\mu}_{\underline{0}} e^{\nu}_{\underline{1}, \underline{2}} = \pm \frac{1}{2} \tensor{\Omega}{^\mu ^\nu _\lambda _\rho} F^{\mathrm{out}}_{\mu \nu} e^{\lambda}_{\underline{2}, \underline{1}} e^{\rho}_{\underline{3}}$, because for a right-handed tetrad we have $\Omega_{\mu \nu \lambda \rho} e^{\mu}_{\underline{0}} e^{\nu}_{\underline{1}} e^{\lambda}_{\underline{2}} e^{\rho}_{\underline{3}} = 1$.}
Therefore, finally, taking $A \to 0$ we find that
\begin{equation}
	F^{\mathrm{out}}_{\mu \nu} e^{\mu}_{\underline{0}} e^{\nu}_{\underline{1}, \underline{2}} = F^{\mathrm{in}}_{\mu \nu} e^{\mu}_{\underline{0}} e^{\nu}_{\underline{1}, \underline{2}} = 0 \, ,
\end{equation}
where we have used that inside the conductor, the electric field vanishes.
This means that the electric field parallel to the conductor's surface has to vanish in any point on this surface, $E_{\parallel} = 0$.

\bigskip

Similarly, we can argue that the magnetic flux through the surface area is conserved.
To see this, let us consider a three-dimensional volume that is tangential to $e_{\underline{1}}$, $e_{\underline{2}}$ and $e_{\underline{3}}$.
Following the same argument as above, we find
\begin{equation}
    f^*F^{\mathcal{B}} \left(e_{\underline{1}},e_{\underline{2}}\right)=F^\mathrm{in}_{\mu\nu}e_{\underline{1}}^\mu e_{\underline{2}}^\nu=F^\mathrm{out}_{\mu\nu}e_{\underline{1}}^\mu e_{\underline{2}}^\nu \, .
\end{equation}
We note that the measured magnetic field orthogonal to the conductor's surface is continuous, but not necessarily constant in time.
Naively, the time dependence only originates from $e_{\underline{1}}$ and $e_{\underline{2}}$, which span a unit area $A$.
One might therefore interpret this relation as the magnetic flux through this unit area being conserved, $B_{\bot} \times A = \mathrm{const}$.
More precisely, we can consider an area $A$ small enough that $f:A \to f(A)$ is a diffeomorphism.
Then consider a second area $A^{\prime}$ (not necessarily containing $x$), such that $f(A^{\prime})=f(A)$ and $f:A^{\prime} \to f(A^{\prime})$ is also a diffeomorphism.
In this scenario, naively, $A$ and $A^{\prime}$ denote the same surface area of the conductor, but at different times.
We then obtain
\begin{equation}
    \int_{A}F=\int_{f(A)}F^{\mathcal{B}}=\int_{A^{\prime}}F \, .
\end{equation}
Therefore, the magnetic flux through the conductor's surface is conserved.

\bigskip

While the above discussion provides conditions on both the parallel component of the electric field and the normal component of the magnetic field, we can use the formalism to consider their respective normal and parallel analogue, respectively.
We begin by defining the current 3-form as $J = \star j_{\mu} \md x^{\mu}$, where $\star$ denotes the Hodge dual.
The inhomogeneous part of Maxwell's equations then reads
\begin{equation}
	\md \star F = J \, .
\end{equation}
Note that, similar to the electric field, the electric polarisation of an ideal conductor vanishes.
Nevertheless they may still feature a nonvanishing magnetisation.
As it is typically small in many materials, we neglect this possibility however.
By Stokes' theorem we can therefore write
\begin{equation}
	\int_V J = \int_V \md  \star F = \oint_{\partial V} \star F \, .
\end{equation}
We again start by considering the spacetime volume $V$ that is tangential to $e_{\underline{0}}$, $e_{\underline{3}}$, and either $e_{\underline{1}}$ or $e_{\underline{2}}$.
Using the volume form, we can evaluate the left hand side of this equation to
\begin{equation}
	\int_V J = \int_V \md V \, \tensor{\Omega}{^\mu _\nu _\rho _\sigma} j_{\mu} e_{\underline{0}}^{\nu} e_{\underline{1},\underline{2}}^{\rho} e_{\underline{3}}^{\sigma}= \pm \int_V \md V \, j_{\mu} e_{\underline{2},\underline{1}}^{\mu} \, ,
\end{equation}
while, in the limit $\epsilon \to 0$, the right hand side can be written as
\begin{equation}
	\oint_{\partial V} \star F = \pm \int_A \md A \, F^{\mathrm{out}}_{\mu \nu} e_{\underline{2}, \underline{1}}^{\mu} e_{\underline{3}}^{\nu}
	\mp \int_A \md A \, F^{\mathrm{in}}_{\mu \nu} e_{\underline{2}, \underline{1}}^{\mu} e_{\underline{3}}^{\nu} \, .
\end{equation}
Following~\cite{Rawson-Harris:1972nfp} and defining the surface current in the point $x$ that is measured by an observer whose coordinates coincide with the tetrad,
\begin{equation}
	I_{\underline{1},\underline{2}} = \lim_{A \to 0} \frac{\lim_{\epsilon \to 0} \int_V \md V \, j_{\mu} e_{\underline{1},\underline{2}}^{\mu}}{\int_A \md A} \, ,
\end{equation}
we finally obtain the relation between the magnetic field parallel to the conductor's surface and the corresponding surface current,
\begin{equation}
	I_{\underline{1},\underline{2}} = F^{\mathrm{out}}_{\mu \nu} e_{\underline{1}, \underline{2}}^{\mu} e_{\underline{3}}^{\nu}
	- F^{\mathrm{in}}_{\mu \nu} e_{\underline{1}, \underline{2}}^{\mu} e_{\underline{3}}^{\nu} \, .
\end{equation}
To find the relation between the surface charge $Q$ and the normal component of the electric field we can proceed in the same manner, but now with a 3-volume that is tangential to $e_{\underline{1}}$, $e_{\underline{2}}$ and $e_{\underline{3}}$, and obtain
\begin{equation}
	Q = \lim_{A \to 0} \frac{\lim_{\epsilon \to 0} \int_V \md V \, j_{\mu} e_{\underline{0}}^{\mu}}{\int_A \md A} = F^{\mathrm{out}}_{\mu \nu} e_{\underline{0}}^{\mu} e_{\underline{3}}^{\nu} \, ,
\end{equation}
where we have used that the electric field inside the conductor vanishes, $F^{\mathrm{in}}_{\mu \nu} e_{\underline{0}}^{\mu} e_{\underline{3}}^{\nu} = 0$.

\section{Details on the Perturbation Scheme}
\label{app:PerturbationScheme}

Here we present a few additional details on the perturbation scheme defined in \cref{sec: electrodynamics and gravitational waves}.
In particular, we give a summary of the transformation laws of the vector field fluctuations that guarantee the invariance of the observable electromagnetic fields under infinitesimal coordinate transformations.

In general, the transformation properties of tensor quantities may differ for different choices of a perturbation scheme.
This is because the indices of the unperturbed tensors are raised and lowered using the metric $g_{\mu \nu}$, while the indices of all tensors in the perturbative expansion are raised and lowered using the flat background metric tensor $\eta_{\mu \nu}$.
Indeed, as we expand to linear order in the metric fluctuations $h_{\mu\nu}$, any other prescription would introduce higher-order perturbations in $h_{\mu\nu}$.
Let us be very explicit and consider a crucial example of our perturbation scheme,
\begin{align}
	g_{\mu \nu} &= \eta_{\mu \nu} + h_{\mu \nu} \, , \\
	F_{\mu \nu} &= \bF_{\mu \nu} + \delta F_{\mu \nu} \, ,
\end{align}
where $\delta F_{\mu \nu}$ is of order $\mathcal{O}(h)$, and $\bF_{\mu \nu}$ is the leading order term in this expansion.
As we have mentioned before, the indices of the unperturbed electromagnetic field strength tensor are raised and lowered using $g_{\mu \nu}$.
On the other hand, both $\bF_{\mu \nu}$ and $\delta F_{\mu \nu}$ are transformed using $\eta_{\mu \nu}$.
This leads to the somewhat curious observation that the perturbative expansion of $F^{\mu \nu}$ will contain additional terms as compared to the perturbative expansion of $F_{\mu \nu}$.
This can, for instance, be seen from the perturbative expansion of the inverse metric which reads
\begin{equation}
	g^{\mu \nu} = \eta^{\mu \nu} - h^{\mu \nu} \, ,
\end{equation}
in turn guaranteeing that $g_{\mu \lambda}g^{\lambda \nu} = \delta_{\mu}^{\nu}$.
In our choice of scheme we obtain
\begin{equation}
	F^{\mu \nu} = g^{\mu \lambda} g^{\nu \rho} F_{\lambda \rho} = \bF^{\mu \nu} + \delta F^{\mu \nu} - \tensor{h}{^\mu _\lambda} \bF^{\lambda \nu} - \tensor{h}{^\nu _\rho} \bF^{\mu \rho} \, ,
\end{equation}
where we have again neglected terms beyond linear order in the metric fluctuation. 
Clearly, this is drastically different from the perturbative expansion of $F_{\mu \nu}$ given above.
In a scheme based on the expansion $F^{\mu \nu} = \bF^{\mu \nu} + \dF^{\mu \nu}$, the equations of motion therefore differ.
For instance, the homogeneous Maxwell's equations acquire additional source terms (see, e.g.,~\cite{Sokolov:2022dej}).
Here, for convenience, we give a complete list of the perturbation scheme that we use in this work
\begin{gather}
	g_{\mu \nu} = \eta_{\mu \nu} + h_{\mu \nu} \, , \enspace e_{\underline{\alpha}}^{\mu} = \overline{e}_{\underline{\alpha}}^{\mu} + \delta e_{\underline{\alpha}}^{\mu} \, , \\
	x^{\mu} = \overline{x}^{\mu} + \delta x^{\mu} \, , \enspace u^{\mu} = \overline{u}^{\mu} + \delta u^{\mu} \, , \\
	F_{\mu \nu} = \bF_{\mu \nu} + \delta F_{\mu \nu} \, , \enspace j^{\mu} = \bJ^{\mu} + \delta j^{\mu} \, .
\end{gather}
We also remark that, among others, the electromagnetic field strength tensor field is a local object.
Therefore, there is a nontrivial interplay between the perturbative expansion of $x^{\mu}$ and its tensor components.
This, in turn, is crucial for the perturbative expansion of the observable electromagnetic fields that we discuss in \cref{sec: electrodynamics and gravitational waves}.
For instance, the coordinate-invariant expression for the electric field~\eqref{eq:Eobs} reads
\begin{equation}
	E_{\underline{a}} = F_{\mu \nu}\left(x(\tau)\right) e_{\underline{a}}^{\mu}(\tau) u^{\nu}(\tau) \, ,
\end{equation}
where we have explicitly indicated all quantities as a function of proper time.
In our choice of perturbation scheme, this expression becomes
\begin{equation}
	E_{\underline{a}} = \left( \bF_{\mu \nu} (x + \delta x) + \delta F_{\mu \nu} (x + \delta x) \right)
	\left( \overline{e}_{\underline{a}}^{\mu} + \delta e_{\underline{a}}^{\mu} \right)
	\left( \overline{u}^{\nu} + \delta u^{\nu} \right) \, .
\end{equation}
Performing a Taylor expansion while only keeping terms up to linear order in the metric fluctuation, we finally obtain
\begin{equation}
	E_{\underline{a}} = 
	\bF_{\mu \nu} \overline{e}_{\underline{a}}^{\mu} \overline{u}^{\nu}
	+ \delta x^{\lambda} \left( \pd_{\lambda} \bF_{\mu \nu} \right) \overline{e}_{\underline{a}}^{\mu} \overline{u}^{\nu}
	+ \delta F_{\mu \nu} \overline{e}_{\underline{a}}^{\mu} \overline{u}^{\nu}
	+ \bF_{\mu \nu} \delta e_{\underline{a}}^{\mu} \overline{u}^{\nu}
	+ \bF_{\mu \nu} \overline{e}_{\underline{a}}^{\mu} \delta u^{\nu} \, ,
\label{eq:EobsApp}
\end{equation}
which is the general expression of the observable electric field in our choice of perturbation scheme.

\subsection*{Invariance under infinitesimal coordinate transformations}

As we have pointed out in the main text, our framework is invariant under infinitesimal coordinate transformations,
\begin{equation}
	x^{\prime \mu} = x^{\mu} + \xi^{\mu}(x) \, .
\label{eq:CoordinateShiftApp}
\end{equation}
Here, the local coordinate shift $\xi^{\mu}$ is considered to be small, $\pd_{\mu} \xi^{\mu} \sim \mathcal{O}(h)$.
Let us discuss this invariance more explicitly.
Under a general coordinate transformation, $x^{\mu} \to x^{\prime \mu}$, any local tensor field $T$ transforms as
\begin{equation}
	\tensor{{T^{\prime}}}{^{\mu_1}^{\mu_2}^{\ldots} _{\nu_1}_{\nu_2}_{\ldots}} \left(x^{\prime}\right) =
	\tensor{T}{^{\rho_1}^{\rho_2}^{\ldots} _{\sigma_1}_{\sigma_2}_{\ldots}} \left(x\right)
	\frac{\partial x^{\prime \mu_1}}{\partial x^{\rho_1}} \frac{\partial x^{\prime \mu_2}}{\partial x^{\rho_2}} \ldots
	\frac{\partial x^{\sigma_1}}{\partial x^{\prime \nu_1}} \frac{\partial x^{\sigma_2}}{\partial x^{\prime \nu_2}} \ldots \, .
\end{equation}
For instance, after an infinitesimal coordinate shift~\eqref{eq:CoordinateShiftApp} the electromagnetic field strength tensor in the new coordinates reads
\begin{equation}
	F^{\prime}_{\mu \nu} \left(x^{\prime}\right) = F_{\rho \sigma} (x)
	\frac{\partial x^{\rho}}{\partial x^{\prime \mu}} \frac{\partial x^{\sigma}}{\partial x^{\prime \nu}}
	= F_{\mu \nu}(x) - F_{\mu \lambda}(x) \pd_{\nu} \xi^{\lambda} - F_{\lambda \nu}(x) \pd_{\mu} \xi^{\lambda} + \ldots \, .
\end{equation}
At the same time, we can perform a Taylor expansion on the left hand side of this equation, and obtain
\begin{equation}
	F^{\prime}_{\mu \nu} \left(x^{\prime}\right) = F^{\prime}_{\mu \nu}(x) + \xi^{\lambda} \pd_{\lambda} F^{\prime}_{\mu \nu} (x) + \ldots \, .
 \label{eq:TransFromArg}
\end{equation}
We can now apply the perturbation scheme to the electromagnetic field strength tensor and read off its local transformation properties at each order in the perturbative expansion.
In general, following this prescription, we can find the transformation properties of all tensor objects in our choice of scheme.
While the leading-order contributions do not transform under infinitesimal coordinate transformations, e.g.
\begin{equation}
	\bF^{\prime}_{\mu \nu} = \bF_{\mu \nu} \, ,
\end{equation}
the linear terms have nontrivial transformation properties.
For convenience, we summarise these here,
\begin{gather}
	h^{\prime}_{\mu \nu} = h_{\mu \nu} - \pd_\mu \xi_\nu - \pd_\nu \xi_\mu \, , \enspace
	\delta e_{\underline{\alpha}}^{\prime \mu} = \delta e_{\underline{\alpha}}^{\mu} + \overline{e}_{\underline{\alpha}}^{\nu} \pd_{\nu} \xi^{\mu} \, , \\
	\delta x^{\prime \mu} = \delta x^{\mu} + \xi^{\mu} \, , \enspace
	\delta u^{\prime \mu} = \delta u^{\mu} + \overline{u}^{\nu} \pd_{\nu} \xi^{\mu} \, , \\
	\dF^{\prime}_{\mu \nu} = \dF_{\mu \nu} - \xi^{\lambda} \pd_{\lambda} \bF_{\mu \nu} - \bF_{\lambda \nu} \pd_\mu \xi^{\lambda} - \bF_{\mu \lambda} \pd_\nu \xi^{\lambda} \, , \enspace
	\dJ^{\prime \mu} = \dJ^{\mu} - \xi^{\lambda} \pd_{\lambda} \bJ^{\mu} + \bJ^{\lambda} \pd_{\lambda} \xi^\mu \, .
\end{gather}
We remark that the perturbations of the tetrad, $\delta e_{\underline{\alpha}}^{\mu}$, transform slightly different compared to the current, $\delta j^{\mu}$.
This is because, unlike the current, it is not a vector field mapping points from the spacetime manifold into its tangent space.
Instead the tetrad is a function that maps proper time, i.e.~a real parameter, into the tangent space for a curve in the spacetime manifold.
The terms arising from \cref{eq:TransFromArg} are therefore omitted.
Using the above transformation laws, it is straightforward to explicitly check that, e.g., the observable electric field~\eqref{eq:EobsApp} is indeed invariant under infinitesimal coordinate transformations, and therefore well defined.

\bigskip

Finally, before we close this discussion, we note that apart from this class of perturbation schemes (i.e.~making a choice of trivially expanding either $F_{\mu\nu}$ or $F^{\mu\nu}$), there are also schemes that rely on the introduction of a field of tetrads.
By integrating the time-like component of this tetrad field, one obtains a threading of spacetime that is typically interpreted as the worldlines of a family of observers (see, e.g., \cite{Sorge:2023nax}).
In particular, this tetrad field allows for a definition of the observable fields by the respective observer going through an arbitrary event $x$, such that $F_{\underline{\alpha\vphantom{\beta}} \underline{\beta}}(x)=F_{\mu\nu}(x)e^{\mu}_{\underline{\alpha\vphantom{\beta}}}(x)e^{\nu}_{\underline{\beta}}(x)$.
The perturbation scheme is then set up in $F_{\underline{\alpha\vphantom{\beta}} \underline{\beta}}$.
The appeal of this procedure is apparent, as the boundary conditions and the observable field become trivial if the family of observers is chosen carefully to represent the mechanical properties of the detector.
While it seems plausible that this procedure leads to the same results as the perturbation scheme laid out here, we leave a thorough investigation of this approach to future work.
We also refer to Ref.~\cite{Hwang:2023jhs} for an intermediate approach, where only a four-velocity field $u(x)$ is introduced and, for instance, $F_{\mu \nu}(x)u^\mu(x)$ is perturbed.
We note, however, that this quantity is not coordinate invariant and hence cannot represent an observable electromagnetic field as is.

\section{Determination of the Mechanical Mode Functions}
\label{app:DeformationCoefficients}

A general ansatz to solve the dynamics of elastic deformations of the hollow sphere, given in \cref{eq:MechanicsEOMSphere}, can be written as
\begin{equation}
	\delta \vec{x} = \left( \nabla \chi + i \nabla \times \vec{L} \phi + i \vec{L} \psi \right) \me^{i \omega t} \, .
\end{equation}
Here, the function $\chi$ characterises longitudinal waves, while both $\phi$ and $\psi$ characterise transverse waves.
These can be parametrised by an expansion in spherical harmonics,
\begin{equation}
	\chi (r, \theta, \varphi) = \sum_{l=0}^{\infty} \sum_{m=-l}^l \left( \chi_{lm} j_l (pr) + \tilde{\chi}_{lm} y_l (pr) \right) Y_l^m (\theta, \varphi) \, ,
\label{eq:ChiSphericalHarmonicsExpansionAppendix}
\end{equation}
and similar for $\phi$ and $\psi$ with $p$ replaced by $q$, whose definitions are given in \cref{eq:P_Q}.
The mode functions $\chi_{lm}$, $\phi_{lm}$ and $\psi_{lm}$ entirely determine the solution $\delta \vec{x}$.
They can be obtained from the boundary condition on the mechanical deformation at the inner and outer wall of the cavity,
\begin{equation}
	\left. \left( \lambda \vec{e}_r \left( \nabla \cdot \delta \vec{x} \right) + 2 \mu \pd_r \delta \vec{x} + \mu \vec{e}_r \times \left(\nabla \times \delta \vec{x} \right) + \mu \vec{y} \right) \right\rvert_{r=R,R+\Delta R} = 0 \, .
\end{equation}
To solve this boundary condition, it is convenient to split it into components that are normal and tangential to the spherical surface.
Clearly, the normal component is given by the projection onto the radial component $\vec{e}_r$.
For the tangential components, proportional to $\vec{e}_{\theta}$ and $\vec{e}_{\varphi}$, we introduce the surface gradient and the angular momentum operator
\begin{equation}
	\nabla_S = \frac{1}{r} \left( \vec{e}_\theta \pd_\theta + \frac{1}{\sin \theta} \vec{e}_{\varphi} \pd_{\varphi} \right) \, , \enspace
	\vec{L} = i \left( \frac{1}{\sin \theta} \vec{e}_{\theta} \pd_{\varphi} - \vec{e}_{\varphi} \pd_{\theta} \right) \, .
\label{eq:SurfaceGradients}
\end{equation}
This is essentially a projection of the gradient onto the spherical surface and its orthogonal equivalent.
For both we can similarly define a surface divergence following the construction of the Laplace-Beltrami operator (see, e.g., \cite{2006OptCo.265...52A}).
In spherical coordinates, the divergence operators are written as
\begin{equation}
	\nabla_S \cdot \vec{A} = \frac{1}{r \sin \theta} \left( \pd_{\theta} \left( \sin \theta A_{\theta} \right) + \pd_{\varphi} A_{\varphi} \right) \, , \enspace
	\vec{L} \cdot \vec{A} = \frac{i}{\sin \theta} \left( \pd_{\varphi} A_{\theta} - \pd_{\theta} \left( \sin \theta A_{\varphi} \right) \right) \, ,
\end{equation}
for some vector field $\vec{A}$.
Using these projections, the boundary condition, schematically denoted by $\vec{M} = 0$, reads
\begin{align}
	\vec{e}_r \cdot \vec{M} &= \left. \left( \lambda \nabla^2 \chi + 2 \mu \pd_r \left( \pd_r \chi - \frac{1}{r} \vec{L}^2 \phi \right) + \mu \vec{e}_r \cdot \vec{y} \right) \right\rvert_{r=R,R+\Delta R} \, , \\
	\nabla_S \cdot \vec{M} &=  \left. \left( -\frac{2\mu}{r} \vec{L}^2 \pd_r \left(\frac{\chi}{r}\right) + \frac{\mu}{r} \vec{L}^2 \left( \pd_r^2 \phi - \frac{2}{r^2} \phi + \frac{1}{r^2} \vec{L}^2 \phi \right) + \mu \nabla_S \cdot \vec{y} \right) \right\rvert_{r=R,R+\Delta R} \, , \\
	\vec{L} \cdot \vec{M} &= \left. \left( i \mu r \vec{L}^2 \pd_r \left( \frac{\psi}{r} \right) + \mu \vec{L} \cdot \vec{y} \right) \right\rvert_{r=R,R+\Delta R} \, .
\end{align}
As the functions $\chi$, $\phi$ and $\psi$ are expanded in terms of spherical harmonics according to \cref{eq:ChiSphericalHarmonicsExpansionAppendix}, we can separately determine the mode functions $\chi_{lm}$, $\phi_{lm}$ and $\psi_{lm}$ by projecting onto the corresponding spherical harmonic, for instance
\begin{equation}
	\int \md \Omega \, \bar{Y}_l^m \chi(r, \theta, \varphi) = \chi_{lm} j_l (pr) + \tilde{\chi}_{lm} y_l (pr) \, ,
\end{equation}
where we have used the orthogonality relation $\int \md \Omega \bar{Y}_l^m Y_{l^{\prime}}^{m^{\prime}} = \delta_{l l^{\prime}} \delta_{m m^{\prime}}$.
Along these lines, the boundary condition can be written as a system of linear equations (see also~\cite{Coccia:1997gy}),
\begin{equation}
	\begin{pmatrix}
		a_{\chi} (R) & a_{\phi} (R) & \tilde{a}_{\chi} (R) & \tilde{a}_{\phi} (R) \\
		b_{\chi} (R) & b_{\phi} (R) & \tilde{b}_{\chi} (R) & \tilde{b}_{\phi} (R) \\
		a_{\chi} (R+\Delta R) & a_{\phi} (R+\Delta R) & \tilde{a}_{\chi} (R+\Delta R) & \tilde{a}_{\phi} (R+\Delta R) \\
		b_{\chi} (R+\Delta R) & b_{\phi} (R+\Delta R) & \tilde{b}_{\chi} (R+\Delta R) & \tilde{b}_{\phi} (R+\Delta R)
	\end{pmatrix}
	\begin{pmatrix}
		\chi_{lm} \\
		\phi_{lm} \\
		\tilde{\chi}_{lm} \\
		\tilde{\phi}_{lm}
	\end{pmatrix}
	= - \mu
	\begin{pmatrix}
		y^{\vec{e}_r}_{lm} (R) \\
		y^{\nabla_S}_{lm} (R) \\
		y^{\vec{e}_r}_{lm} (R+\Delta R) \\
		y^{\nabla_S}_{lm} (R+\Delta R)
	\end{pmatrix} \, ,
\end{equation}
as well as
\begin{equation}
	\begin{pmatrix}
		c_{\psi} (R) & \tilde{c}_{\psi} (R) \\
		c_{\psi} (R + \Delta R) & \tilde{c}_{\psi} (R + \Delta R)
	\end{pmatrix}
	\begin{pmatrix}
		\psi_{lm} \\
		\tilde{\psi}_{lm}
	\end{pmatrix}
	= - \mu
	\begin{pmatrix}
		y^{\vec{L}}_{lm} (R) \\
		y^{\vec{L}}_{lm} (R + \Delta R)
	\end{pmatrix} \, .
\end{equation}
Here, the inhomogeneous terms on the right hand side are defined as
\begin{equation}
	y^{\vec{e}_r}_{lm} = \int \md \Omega \, \bar{Y}_l^m \vec{e}_r \cdot \vec{y} \, , \enspace
	y^{\nabla_S}_{lm} = \int \md \Omega \, \bar{Y}_l^m \nabla_S \cdot \vec{y} \, , \enspace
	y^{\vec{L}}_{lm} = \int \md \Omega \, \bar{Y}_l^m \vec{L} \cdot \vec{y} \, .
\end{equation}
Furthermore, we have introduced the coefficients
\begin{gather}
	a_{\chi} = - \lambda p^2 \beta_0 + 2 \mu \beta_2 \, , \enspace
	a_{\phi} = - 2 \mu l (l+1) \beta_3 \, , \enspace
	c_{\psi} = i \mu l (l+1) r \beta_3 \, , \enspace \\
	b_{\chi} = - 2 \mu \frac{l (l+1)}{r} \beta_3 \, , \enspace
	b_{\phi} = \mu \frac{l (l+1)}{r} \left( \beta_2 - \frac{l (l+1) - 2}{r^2} \beta_0 \right) \, ,
\end{gather}
together with the schematic abbreviations
\begin{equation}
	\beta_0 = j_l(k r) \, , \enspace
	\beta_1 = \pd_r j_l(kr) \, , \enspace
	\beta_2 = \pd_r^2 j_l(kr) \, , \enspace
	\beta_3 = \pd_r \left( \frac{j_l (kr)}{r} \right) \, .
\end{equation}
More precisely, for all terms involving $\chi$ we have $k = p$, while for all terms involving $\phi$ and $\psi$ we have to set $k = q$.
In addition, for the counterparts, such as $\tilde{a}_{\chi}$, the spherical Bessel functions of the first kind, $j_l$, are replaced by the spherical Bessel functions of the second kind, $y_l$.
We carefully remark that our definitions of $\beta_i$ may differ from the ones used in~\cite{Coccia:1997gy}.

Solving the system of linear equations for each combination of integers $l$ and $m$, will entirely determine the mechanical mode functions $\chi_{lm}$, $\phi_{lm}$ and $\psi_{lm}$.
Indeed, the inhomogeneous terms $y^{\vec{e}_r}_{lm}$, $y^{\nabla_S}_{lm}$ and $y^{\vec{L}}_{lm}$, all vanish for $m \neq 2$, such that an incoming GW can only excite modes of this order.

\section{Determination of the Electromagnetic Mode Functions}
\label{app:ElectromagneticCoefficients}

Similar to the mechanical mode functions presented in \cref{app:DeformationCoefficients}, we can determine the electromagnetic mode functions of the boundary contributions to the GW-induced electric and magnetic fields,
\begin{align}
	\vec{E}_{\mathrm{bnd}} &= i \nabla \times \vec{L} \zeta + i \vec{L} \eta \, \\
	\vec{B}_{\mathrm{bnd}} &= - \omega \vec{L} \zeta - \frac{1}{\omega} \nabla \times \vec{L} \eta \, .
\end{align}
Again, we perform an expansion in spherical harmonics,
\begin{equation}
	\zeta (r, \theta, \varphi) = \sum_{l=0}^{\infty} \sum_{m=-l}^{l} \zeta_{lm} j_l (\omega r) Y_l^m (\theta, \varphi) \, ,
\end{equation}
and similarly for $\eta$, where the mode functions $\zeta_{lm}$ and $\eta_{lm}$ are determined by the boundary condition
\begin{equation}
	\left. \left( \vec{E}_{\mathrm{bnd}} + \vec{E}_{\mathrm{blk}} + i \omega \delta \vec{x} \times \bvB \right) \right\rvert_{\parallel, r=R} = 0 \, .
\end{equation}
Here, the subscript $\parallel$ indicates that only the components tangential to the unperturbed surface of the sphere are to be considered.
Similar to the previous section, we can project the above boundary condition using the operators $\nabla_S$ and $\vec{L}$ defined in \cref{eq:SurfaceGradients} before extracting the coefficients of the expansion.
We find
\begin{align}
	\zeta_{lm} &= \left. - \frac{r^2}{l (l+1) \pd_r \left(r j_l(\omega r)\right)}  \int \md \Omega \, \bar{Y}_l^m \left[ \nabla_S \cdot \vec{E}_{\mathrm{blk}} + i \omega \nabla_S \cdot \left(\delta \vec{x} \times \bvB \right) \right] \right\rvert_{r=R} \, , \\
	\eta_{lm} &= \left. \frac{1}{l (l+1) j_l(\omega r)}  \int \md \Omega \, \bar{Y}_l^m \left[ i \vec{L} \cdot \vec{E}_{\mathrm{blk}} - \omega \vec{L} \cdot \left(\delta \vec{x} \times \bvB \right) \right] \right\rvert_{r=R} \, .
\end{align}
In our example, these integrals can be performed analytically.
To do this, we expand the integrand itself in spherical harmonics and then express the result in terms of Wigner $3j$-symbols,
\begin{equation}
	\int \md \Omega \, Y_{l_1}^{m_1} Y_{l_2}^{m_2} Y_{l_3}^{m_3} = \sqrt{\frac{\left(2 l_1 + 1\right) \left(2 l_2 + 1\right) \left(2 l_3 + 1\right)}{4\pi}}
	\begin{pmatrix}
		l_1 & l_2 & l_3 \\
		0 & 0 & 0
	\end{pmatrix}
	\begin{pmatrix}
		l_1 & l_2 & l_3 \\
		m_1 & m_2 & m_3
	\end{pmatrix} \, .
\end{equation}
On a technical note, we remark that to evaluate the cross- and scalar products involving the background magnetic field $\bvB$, it is convenient to express the latter in a spherical basis with respect to spherical coordinates,
\begin{equation}
	\bvB = \sum_{s=\left\{ 0, \pm \right\}} \tilde{B}^s \vec{n}^s \, ,
\end{equation}
with the basis vectors
\begin{equation}
	\vec{n}^{\pm} = \mp \frac{1}{\sqrt{2}} \left( \vec{e}_{\theta} \mp i \vec{e}_{\varphi} \right) \, , \enspace
	\vec{n}^0 = \vec{e}_r \, .
\end{equation}
In this basis, the coefficients $\tilde{B}$ and $\bB$ are related by spin-weighted spherical harmonics~\cite{Scanio:1977, Newman:1966ub},
\begin{equation}
	\tilde{B}^s = \sqrt{\frac{4\pi}{3}} \sum_{m=-1}^{1} \bB^m {}_s Y_1^m(\theta, \varphi) \, .
\end{equation}
Naively, the spin-weighted spherical harmonics are a convenient choice to express the background magnetic field because the tangential-projection operators $\nabla_S$ and $\vec{L}$ can act as ladder operators that raise or lower the spin weight $s$.
More precisely, in this basis they read
\begin{equation}
	\vec{n}^{\pm} \nabla_S f = \frac{1}{\sqrt{2}r}
	\begin{cases}
		\bar{\eth} f \\
		- \eth f
	\end{cases} \, , \enspace
	\vec{n}^{\pm} \vec{L} f = -\frac{1}{\sqrt{2}}
	\begin{cases}
		\bar{\eth} f \\
		\eth f
	\end{cases} \, ,
\end{equation}
where we have defined
\begin{equation}
	\eth = -\left( \pd_{\theta} + \frac{i}{\sin \theta} \pd_{\varphi} \right) \, .
\end{equation}
This operator and its conjugate then act as ladder operators for the spin-weighted spherical harmonics, e.g.~for spin weights $s=0$,
\begin{equation}
	\eth {}_0 Y_l^m = \sqrt{l(l+1)} {}_1 Y_l^m \, , \enspace
	\bar{\eth} {}_0 Y_l^m = -\sqrt{l(l+1)} {}_{-1} Y_l^m \, .
\end{equation}
That said, it is straightforward (but admittedly tedious) to expand the integrand in a spherical basis and perform the integration analytically to obtain the electromagnetic mode functions.
Furthermore, the spin-weighted spherical harmonics enable a closed-form expression for the product $\vec{d} \cdot \vec{B}_{\mathrm{bnd}}$ appearing in \cref{eq:BobsExplicit}, if one similarly introduces the expansion coefficients $\tilde{d}^s$ analogous to the background magnetic field.

\section{The Spherical Cavity in FN coordinates}
\label{app:SphericalCavityFN}

In general, the observable electromagnetic fields in the spherical cavity are coordinate invariant.
If we, however, are interested in the regime $\omega R \ll 1$, we can approximate the expected signal in FN coordinates.
In practice, this means that we need to determine the mechanical deformations as well as the electromagnetic excitations of the spherical cavity due to the incoming GW.

The mechanical deformations are given by the general solution to \cref{eq:MechanicsEOMFN}, which we obtain by splitting the solution into a bulk and a boundary contribution, $\delta \vec{x} = \delta \vec{x}_{\mathrm{blk}} + \delta \vec{x}_{\mathrm{bnd}}$, where the bulk contribution is an arbitrary solution to the inhomogeneous equations of motion.
It is given by
\begin{equation}
	\delta x_{\mathrm{blk}}^i = \frac{1}{2} \tensor{{\htt}}{^i _j}(t) x^j \, .
\end{equation}
In contrast to the bulk solution, the boundary contribution $\delta \vec{x}_{\mathrm{bnd}}$ guarantees that the correct boundary condition~\eqref{eq:MechanicsBC} is fulfilled, and can be straightforwardly obtained by identically following the procedure presented in \cref{app:DeformationCoefficients}.

To find the electromagnetic modes of the spherical cavity in FN coordinates, we can proceed similarly to the calculation in the TT frame presented in \cref{sec:ElectromagneticExcitations}.
More precisely, we again split the solution into a bulk and a boundary contribution, $\delta F_{i0} = E_i^{\mathrm{blk}} + E_i^{\mathrm{bnd}}$ and $\delta F_{ij} = \epsilon_{ijk} ( B_k^{\mathrm{blk}} + B_k^{\mathrm{bnd}} )$, respectively.
Here, the bulk field is an arbitrary solution to the inhomogeneous Maxwell's equations now involving the effective current in FN coordinates, $\vec{j}_{\mathrm{eff}}^{\mathrm{FN}}$, given by \cref{eq:jeff} in the FN frame.
If, in the long-wavelength limit, $\omega R \ll 1$, we neglect contributions to the effective current of orders higher than $\mathcal{O} ( \omega^2 R^2 )$, the solution to Maxwell's equations in the bulk is given by
\begin{align}
	\vec{B}_{\mathrm{blk}}^{\mathrm{FN}} &= -\frac{1}{\omega^2} \nabla \times \vec{j}_{\mathrm{eff}}^{\mathrm{FN}} \, , \\
	\vec{E}_{\mathrm{blk}}^{\mathrm{FN}} &= -\frac{i}{\omega} \left( \nabla \times \vec{B}_{\mathrm{blk}}^{\mathrm{FN}} - \vec{j}_{\mathrm{eff}}^{\mathrm{FN}} \right) \, ,
\end{align}
while the boundary contribution is similar to the ansatz given in \cref{app:ElectromagneticCoefficients}, with the corresponding mode functions
\begin{align}
	\zeta^{\mathrm{FN}}_{lm} &= \left. - \frac{r^2}{l (l+1) \pd_r \left(r j_l(\omega r)\right)}  \int \md \Omega \, \bar{Y}_l^m \left[ \nabla_S \cdot \vec{E}^{\mathrm{FN}}_{\mathrm{blk}} + i \omega \nabla_S \cdot \left(\delta \vec{x}^{\mathrm{FN}} \times \bvB \right) \right] \right\rvert_{r=R} \, , \\
	\eta^{\mathrm{FN}}_{lm} &= \left. \frac{1}{l (l+1) j_l(\omega r)}  \int \md \Omega \, \bar{Y}_l^m \left[ i \vec{L} \cdot \vec{E}^{\mathrm{FN}}_{\mathrm{blk}} - \omega \vec{L} \cdot \left(\delta \vec{x}^{\mathrm{FN}} \times \bvB \right) \right] \right\rvert_{r=R} \, .
\end{align}

Similarly, the fluctuations of the observer's tetrad are parametrised in the FN frame, i.e.~the proper detector frame.
In this scenario, they satisfy
\begin{equation}
	\frac{\md}{\md t} \delta e_{\underline{a}}^{i} + \Gamma^i_{0j} \overline{e}^j_{\underline{a}} = 0 \, ,
\end{equation}
where the Christoffel symbol is now given in FN coordinates.
In the long-wavelength approximation, the solution to the equations of motion then reads
\begin{equation}
	\delta e_{\underline{a}}^i = \frac{i}{2} \omega x^k \left( \htt_{jk} \overline{e}_{\underline{a}}^j \delta_3^i - \tensor{{\htt}}{^{i}_{k}} \overline{e}_{\underline{a}}^3 \right) \, .
\end{equation}
Using that $\vec{\overline{e}}_{\underline{1}} \times \vec{\overline{e}}_{\underline{2}} = \vec{d}$, we can finally write
\begin{equation}
	\left( \delta \vec{e}_{\underline{1}} \times \vec{\overline{e}}_{\underline{2}}
	+ \vec{\overline{e}}_{\underline{1}} \times \delta \vec{e}_{\underline{2}} \right) \cdot \bvB
	= \frac{i}{2 \sqrt{2}} \omega r \sin \theta \left[ \left( \bB^+ d^0 - \bB^0 d^+ \right) h^- \me^{-i \varphi} - \left(\bB^- d^0 - \bB^0 d^- \right) h^+ \me^{i \varphi} \right] \, ,
\end{equation}
which will enter the observable electromagnetic fields, approximated in FN coordinates.
We finally remark that the overall contribution from the effective current $\vec{j}_{\mathrm{eff}}^{\mathrm{FN}}$ parametrically scales as $\mathcal{O} \left(\omega^2 R^2\right)$.
Terms of this order contributing to the tetrad evolution, however, have been neglected as they are subleading.
It is therefore consistent to set $\vec{j}_{\mathrm{eff}}^{\mathrm{FN}}=\vec{E}_{\mathrm{blk}}^{\mathrm{FN}}=\vec{B}_{\mathrm{blk}}^{\mathrm{FN}}=0$.

\bibliographystyle{inspire}
\bibliography{refs, refs_noninspire}

\end{document}